\documentclass[referee]{raa_rb}            % referee version: for submission

\usepackage{graphicx,times}             %for PS/EPS graphics inclusion, new
\usepackage{natbib}
\usepackage{amssymb,amsmath}
\usepackage[utf8]{inputenc}
\usepackage{mathrsfs}

\bibpunct{(}{)}{;}{a}{}{,}

\usepackage[pagebackref=true]{hyperref}
\usepackage{enumitem} 
\usepackage{siunitx}
\begin{document}

  \title{Mock Observations for the CSST Mission: Multi-Channel Imager--Instrument Simulation}
%   \subtitle{I. Place Your Subtitle Here}

   \volnopage{Vol.0 (20xx) No.0, 000--000}      %%preserved for Editor. DOn't remove!
   \setcounter{page}{1}          %%starting page, preserved for Editor. DOn't remove!

   \author{Zhao-Jun Yan %(颜召军)
      \inst{1}
   \and Huan-Yuan Shan %(陕欢源)
      \inst{1}
   \and Zhen-Ya Zheng %(郑振亚) %% Put your Chinese name in "( )" if you like. Note to open line 11 "\usepackage[UTF8]{ctex}"
      \inst{1}
   \and Xi-Yan Peng %(彭喜衍) %% Put your Chinese name in "( )" if you like. Note to open line 11 "\usepackage[UTF8]{ctex}"
      \inst{1}
   \and Zhao-Xiang Qi %(齐朝祥)
      \inst{1}
   \and Chun Xu %(许春) %% Put your Chinese name in "( )" if you like. Note to open line 11 "\usepackage[UTF8]{ctex}"
      \inst{1}
   \and Lin Lin %(林琳)
      \inst{1}
    \and Xin-Rong Wen %(文新荣) %% Put your Chinese name in "( )" if you like. Note to open line 11 "\usepackage[UTF8]{ctex}"
      \inst{1}
    \and Chun-Yan Jiang %(姜春燕)
      \inst{1}
    \and Li-Xin Zheng %(郑立新)
      \inst{1}
    \and Jing Zhong %(钟靖)
      \inst{1}
    \and Fang-Ting Yuan %(袁方婷)
      \inst{1}
    \and Zhen-Lei Chen %(陈震雷)
      \inst{1}
    \and Wei Chen%(陈维)
      \inst{1}
    \and Mao-Chun Wu%(吴茂春) 
      \inst{1}
    \and Zhen-Sen Fu%(傅震森)
       \inst{1}
    \and Ke-Xin Li%(李可新)
       \inst{1} 
    \and Lin Nie%(聂麟)
       \inst{1} 
    \and Chao Liu %(刘超) %% Put your Chinese name in "( )" if you like. Note to open line 11 "\usepackage[UTF8]{ctex}"
       \inst{2}
    \and Nan Li %(李楠) %% Put your Chinese name in "( )" if you like. Note to open line 11 "\usepackage[UTF8]{ctex}"
       \inst{2}
     \and Qiao Wang %(王乔) %% Put your Chinese name in "( )" if you like. Note to open line 11 "\usepackage[UTF8]{ctex}"
       \inst{2} 
    \and Zi-Huang Cao %(曹子皇) %% Put your Chinese name in "( )" if you like. Note to open line 11 "\usepackage[UTF8]{ctex}"
       \inst{2}
    \and Shuai Feng %(冯帅) %% Put your Chinese name in "( )" if you like. Note to open line 11 "\usepackage[UTF8]{ctex}"
       \inst{3}
    \and Guo-Liang Li %(李国亮) %% Put your Chinese name in "( )" if you like. Note to open line 11 "\usepackage[UTF8]{ctex}"
        \inst{4}
    \and Lei Wang%(王蕾)
       \inst{4} 
    \and Cheng-Liang Wei  %(韦成亮) %% Put your Chinese name in "( )" if you like. Note to open line 11 "\usepackage[UTF8]{ctex}"
        \inst{4}
    \and Xiao-Bo Li %(李晓波) %% Put your Chinese name in "( )" if you like. Note to open line 11 "\usepackage[UTF8]{ctex}"
        \inst{5}
    \and Zhang Ban %(班章) %% Put your Chinese name in "( )" if you like. Note to open line 11 "\usepackage[UTF8]{ctex}"
         \inst{5}
    \and Xun Yang  %(杨勋) %% Put your Chinese name in "( )" if you like. Note to open line 11 "\usepackage[UTF8]{ctex}"
         \inst{5}
    \and Yu-Xi Jiang %(姜禹希) %% Put your Chinese name in "( )" if you like. Note to open line 11 "\usepackage[UTF8]{ctex}"
         \inst{5}
    \and De-Zi Liu%(刘德子)
       \inst{6}
    \and Yong-He Chen%(陈永和)
       \inst{7}
    \and Xiao-Hua Liu%(刘晓华)
       \inst{7} 
    \and Fang Xu%(徐放)
       \inst{7} 
    \and Xue Cheng%(程雪)
       \inst{7} 
    \and Yue Su%(苏越)
       \inst{7} 
    \and Tong-Fang Duan%(段童方)
       \inst{7} 
    \and Chao Qi%(齐超)
       \inst{7} 
    \and Na Li%(李娜)
       \inst{7} 
    \and Geng Zheng%(郑庚)
       \inst{7} 
    \and Chong Ma%(马冲)
       \inst{7} 
    \and Jing Tang %(汤静) %% Put your Chinese name in "
        \inst{2}
    \and Ran Li %(李然) %% Put your Chinese name in "
        \inst{8}
    }
   \institute{Shanghai Astronomical Observatories, Chinese Academy of Sciences,
             Shanghai 200030, China; {\it zhaojunyan@shao.ac.cn, zhengzy@shao.ac.cn, hyshan@shao.ac.cn}\\
        \and
             National Astronomical Observatories, Chinese Academy of Sciences, Beijing 100101, China\\
        \and
             College of Physics, Hebei Key Laboratory of Photophysics Research and Application, Hebei Normal University, Shijiazhuang 050024, China.\\
        \and
             Purple Mountain Observatory, Chinese Academy of Sciences, Nanjing 210023, China.\\
        \and
             Space Optics Department, Changchun Institute of Optics, Fine Mechanics and Physics, Chinese Academy of Sciences, Changchun 130033, China.\\
        \and
            South-Western Institute for Astronomy Research, Yunnan University, Kunming, Yunnan, 650500, China\\
        \and
             Shanghai Institute of Technical Physics, Chinese Academy of Sciences, Shanghai, 200083, China.\\
        \and
             School of Physics and Astronomy, Beijing Normal University,  Beijing 100875, China.\\
}

\vs\no
   {\small Received 20xx month day; accepted 20xx month day}

\abstract{The Chinese Space Station Survey Telescope (CSST), a two-meter aperture astronomical space telescope under China’s manned space program, is equipped with multiple back-end scientific instruments. As an astronomical precision measurement module of the CSST, the Multi-Channel Imager (MCI) can cover a wide wavelength range from ultraviolet to near-infrared with three-color simultaneous high-precision photometry and imaging, which meets the scientific requirements for various fields. The diverse scientific objectives of MCI require not only a robust airborne platform, advanced optical systems, and observing facilities but also comprehensive software support for scientific operations and research. To this end, it is essential to develop realistic observational simulation software to thoroughly evaluate the MCI data stream and provide calibration tools for future scientific investigations. The MCI instrument simulation software will serve as a foundation for the development of the MCI data processing pipeline and will facilitate improvements in both hardware and software, as well as in the observational operation strategy, in alignment with the mission's scientific goals. In conclusion, we present a comprehensive overview of the MCI instrument simulation and some corresponding performances of the MCI data processing pipeline.
\keywords{telescopes ---  methods: numerical ---  techniques: image processing}
}

   \authorrunning{Zhaojun Yan et al.}            %author_head in even pages
   \titlerunning{Mock Observations for the CSST Mission: Multi-Channel Imager--Instrument simulation}  % title_head in odd pages

   \maketitle
%% The author head (on even pages) and the title head (on odd pages) will be
%% automatically extracted from \author{} and \title{}. Whenever the title is too long,
%% you will be asked to supply a shorter one by inserting either \authorrunning{} or
%% \titlerunning{} before \maketitle. Anyway, you can specify your own heads.
%%
%%
%% Note: In the following text body of your manuscript, please note several differences from
%%       other major journals:
%% (1) \subsection{Please Capitalize the First Letter of Each Notional Word in Subsection Title}
%% (2) Please Capitalize the First Letter of Each Notional Word in all tables' captions

%
%________________________________________________ sections below
%
\section{Introduction}           %% first-level sections will be auto-capitalized
\label{sect:intro}

The Chinese Space Station Survey Telescope (CSST), a two-meter aperture space telescope under China’s manned space program, uniquely combines a wide field of view with high-resolution imaging capabilities \citep{2011Zhan, 2021Zhan}. Equipped with five advanced scientific instruments—a multi-band imaging and slitless spectroscopy survey camera (SC), a multi-channel imager (MCI), an integral field spectrograph (IFS), a cool planet imaging coronagraph (CPIC), and a terahertz spectrometer (TS)—the CSST will conduct simultaneous deep multicolor imaging and slitless spectral surveys across large sky areas. This powerful combination of instruments, working at unprecedented spatial resolution, will enable comprehensive studies of celestial objects. This is expected to contribute to breakthroughs in key problems in astronomy and fundamental physics, including dark matter, dark energy, galaxy formation and evolution, early cosmic chemical enrichment, etc.

The CSST-MCI, a pivotal component of the astronomical back-end module, is capable of conducting simultaneous three-channel photometric observations across the ultraviolet (UV), visible (VIS), and near-infrared (NIR) spectra. The design of each of the three channels of the MCI, which are equipped with multiple filters of varying performance, is tailored to meet a diverse array of scientific requirements, ranging from the study of the solar system to the exploration of the distant universe. Each detector utilizes a 9k$\times$ 9k CCD, which guarantees a substantial field of view and optimizes the spatial sampling rate through focusing, thereby attaining a field of view of $7.5'$ by $7.5'$ and a pixel resolution of 0.05\arcsec. The primary objective of the MCI is to provide high-precision flux standard star catalogs for the main survey, thereby reducing systematic errors through accurate photometry. This, in turn, provides the necessary guarantee for the main survey to achieve its high-precision scientific goals. The other core scientific mission of the MCI is to conduct ultra-deep UV-VIS field observations. These observations have the potential to yield unique and highly competitive scientific benefits that extend beyond the primary survey objectives. MCI will result in ultra-deep exposure depths in the ultraviolet and visible bands, especially in the UV band, which has never been achieved before. For a more comprehensive overview of the MCI, please refer to Zheng et al. (in Prep.).

To achieve the scientific goals of CSST-MCI, it is essential to have a highly efficient space-based observatory, as well as optical and observing facilities. In addition, the provision of software support for scientific operations and research through a comprehensive scientific data system is also necessary. The primary objective of the CSST scientific data system is to generate and process raw images, spectral data and other products that can be used by astronomers. To evaluate the MCI data stream comprehensively and provide calibration tools for future scientific exploration, establishing a comprehensive set of realistic observational simulation software is imperative. The instrument simulation for the CSST-MCI will serve as a basis for the development of the MCI data pipeline and will improve the development of the software and hardware for the MCI on the CSST, as well as the observational operation plan, in line with the achievements of the scientific objectives.

In this paper, we introduce the instrument simulating philosophy of the CSST-MCI, as well as its detailed implementation process. The organization of this paper is as follows. Section 2 delineates the comprehensive optical architecture of the CSST and MCI. Section 3 provides a detailed exposition of the engineering simulation of the CSST main optical system. Section 4 provides a comprehensive description of the detailed simulation flow of CSST-MCI, encompassing the point spread function (PSF) of the optical system, the input star and galaxy catalogs, the sky background, and various instrumental effects. The concluding section synthesizes key findings and discusses implications for future space-based imaging instrumentation.

\section{Optical system introduction}
\label{sect:Obs}
In this section, an exposition of the primary telescope optics, in conjunction with the constituent components of the MCI optics, is provided.

\subsection{The main telescope}
\begin{figure}
   \centering
 \includegraphics[width=\textwidth, angle=0, scale=0.99]{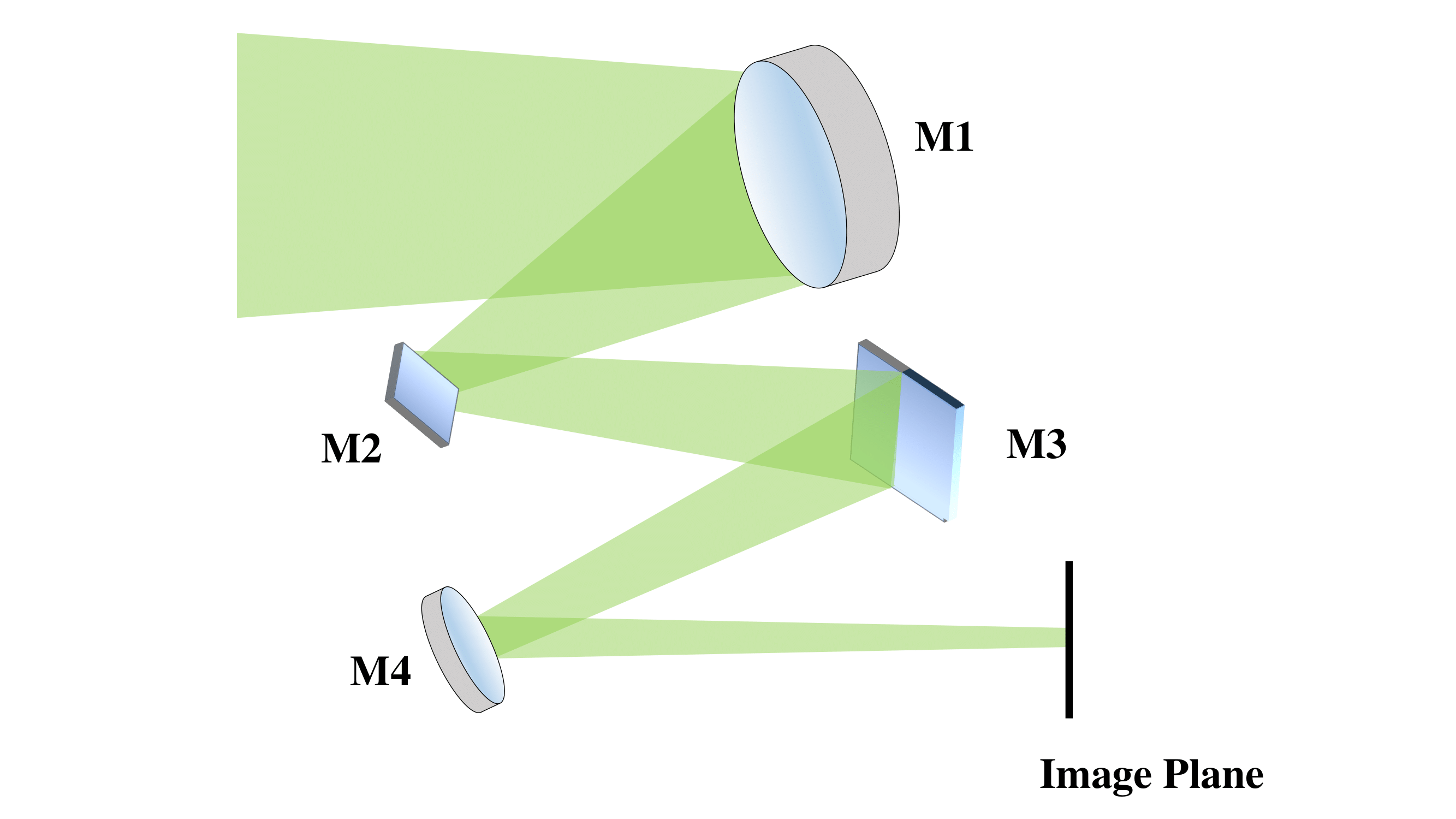}
   \caption{Schematic diagram of the optical path of the optical system of the CSST.}
   \label{Fig1}
\end{figure}

The main telescope optical system of the CSST adopts the structure of the COOK off-axis three-mirror optical system to achieve off-axis unobstructed imaging, as shown in Figure~\ref{Fig1}. The system consists of four optical mirrors M1--M4, of which M1 and M2 use quadratic surfaces, M3 uses a free-form surface, and M4 uses a plane surface. The clear aperture is 2 m and the focal length is approximately 28 m. The field of view is 1.72 square degrees.

\subsection{MCI optical system}

\begin{figure}
   \centering
   \includegraphics[width=\textwidth, angle=0, scale=1.0]{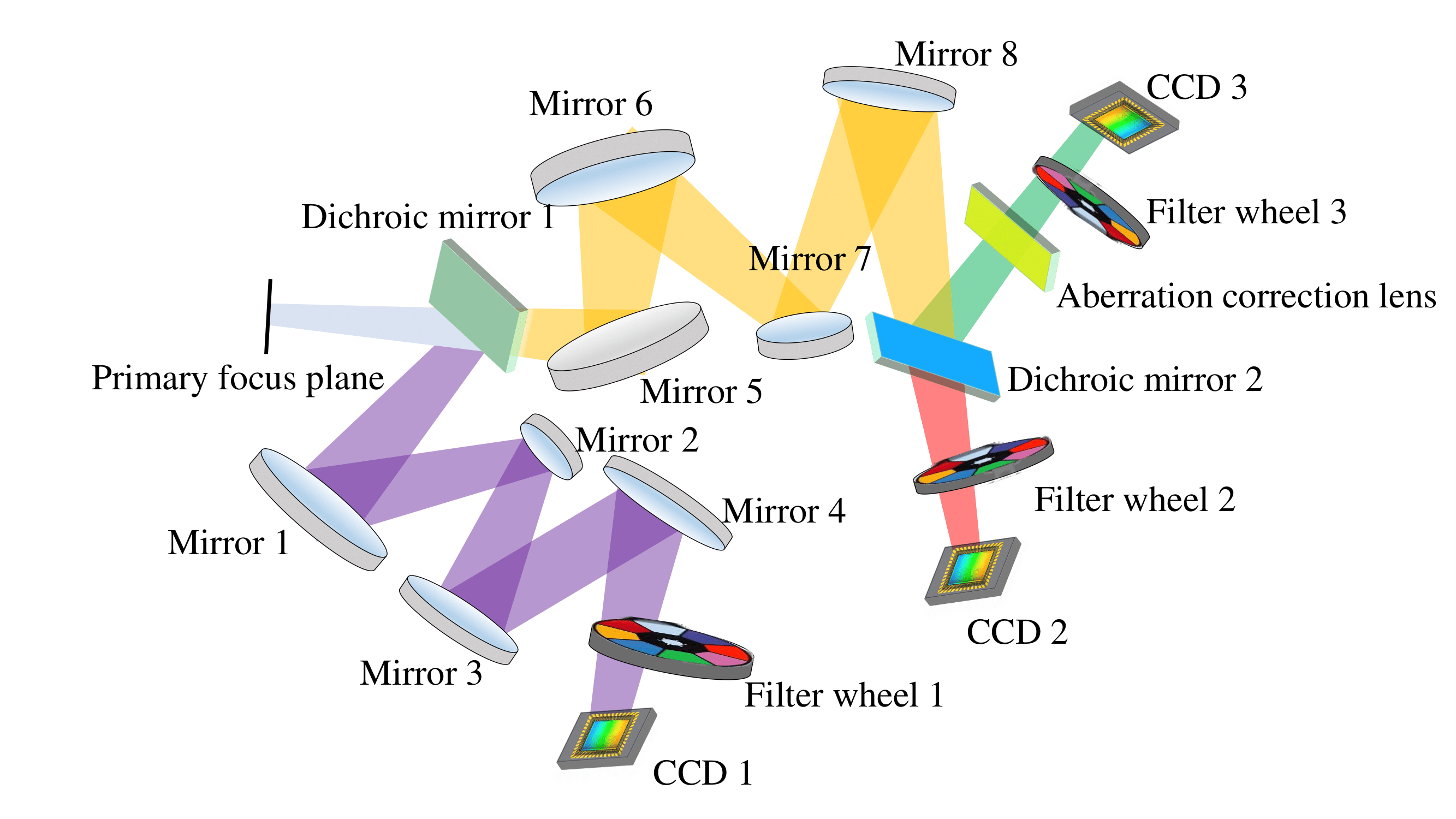}
   \caption{Schematic diagram of the optical system of the MCI.}
   \label{Fig2}
\end{figure}

A schematic representation of the MCI optical system is depicted in Figure~\ref{Fig2}. Dichroic mirror 1 splits the main telescope’s focal-plane beam into two paths: a reflected ultraviolet beam and a transmitted visible beam. The reflected beam sequentially traverses mirrors 1 through 4, then passes through the optical filter mounted on filter wheel 1, and is ultimately captured by CCD 1, serving as the ultraviolet band detection unit. Another beam is the transmitted beam of the visible band. After passing through mirrors 5 $\thicksim$ 8 in sequence, the beam is split into two different beams by dichroic mirror 2. One beam is the reflected beam of the visible band, which passes through the aberration correction lens and the filter in filter wheel 3 and is finally detected by CCD 3. The other beam, transmitted in the near-infrared band, reaches detector 2 after passing through filter wheel 2. The main function of the MCI optical system is to split the main telescope beam into three channels. The second function is to vary the focal length of the system. The focal length of the main telescope system is 28 m, whereas the equivalent focal length of the MCI optical system is 42 m.

\section{Engineering simulation of the CSST}
\label{sect:data}

The CSST primary and MCI optical systems are designed to mitigate various types of wavefront errors (WFEs). Mid-spatial frequency aberrations scatter light at small angles around the PSF centroid, reducing image contrast, whereas high-spatial frequency errors cause wider-angle scattering, lowering overall system throughput. Each WFE regime affects the MCI’s scientific performance differently. To support the CSST scientific data system, we conducted a comprehensive engineering simulation of the telescope, calculating wavefront aberrations and PSFs under the influence of multiple error sources. These include optical design imperfections, mirror fabrication errors, assembly misalignments, gravitational effects, and thermal deformations. The design-optimized optical system achieves an average aberration better than $\lambda/30$ across the field of view and an $80\%$ encircled energy radius below 0.1\arcsec, reaching diffraction-limited performance. Based on current optical fabrication capabilities, mirror surface errors were estimated via numerical simulation. The test data, decomposed into low- and mid-frequency components, were analyzed to characterize the errors's frequencies. The low-frequency components were fitted via Zernike polynomials, whereas a function-based simulator generated the mid- to high-frequency components. Taken together, these factors constitute the total optical fabrication error, resulting in a root mean square (RMS) wavefront error of 0.013$\lambda$. Upon entering orbit, gravitational changes deform the telescope structure, altering the mirror positions and surface shapes. The secondary and three-in-one mirrors are mounted on the front frame, while the primary and associated mirrors are attached to the main substrate. Structural deformation from gravitational shifts affects the entire mirror assembly. Finite element analysis (FEA) was used to quantify these displacements and surface deformations. In orbit, thermal loads from external flux and internal heat sources further deform the opto-mechanical structure, degrading image quality. To simulate these effects under extreme temperature conditions, FEA was again applied to assess the mirror displacements and surface changes. Additionally, two dynamic error sources were considered: high-frequency optical axis jitter error caused by microvibrations, and low-frequency drift from residual image stabilization errors. Both were modeled and incorporated into the simulation. Further technical details are provided in a companion paper (Ban et al., in preparation, 2025).

\section{MCI instrument simulation}

\begin{figure}
   \centering
 \includegraphics[width=\textwidth, angle=0, scale=0.99]{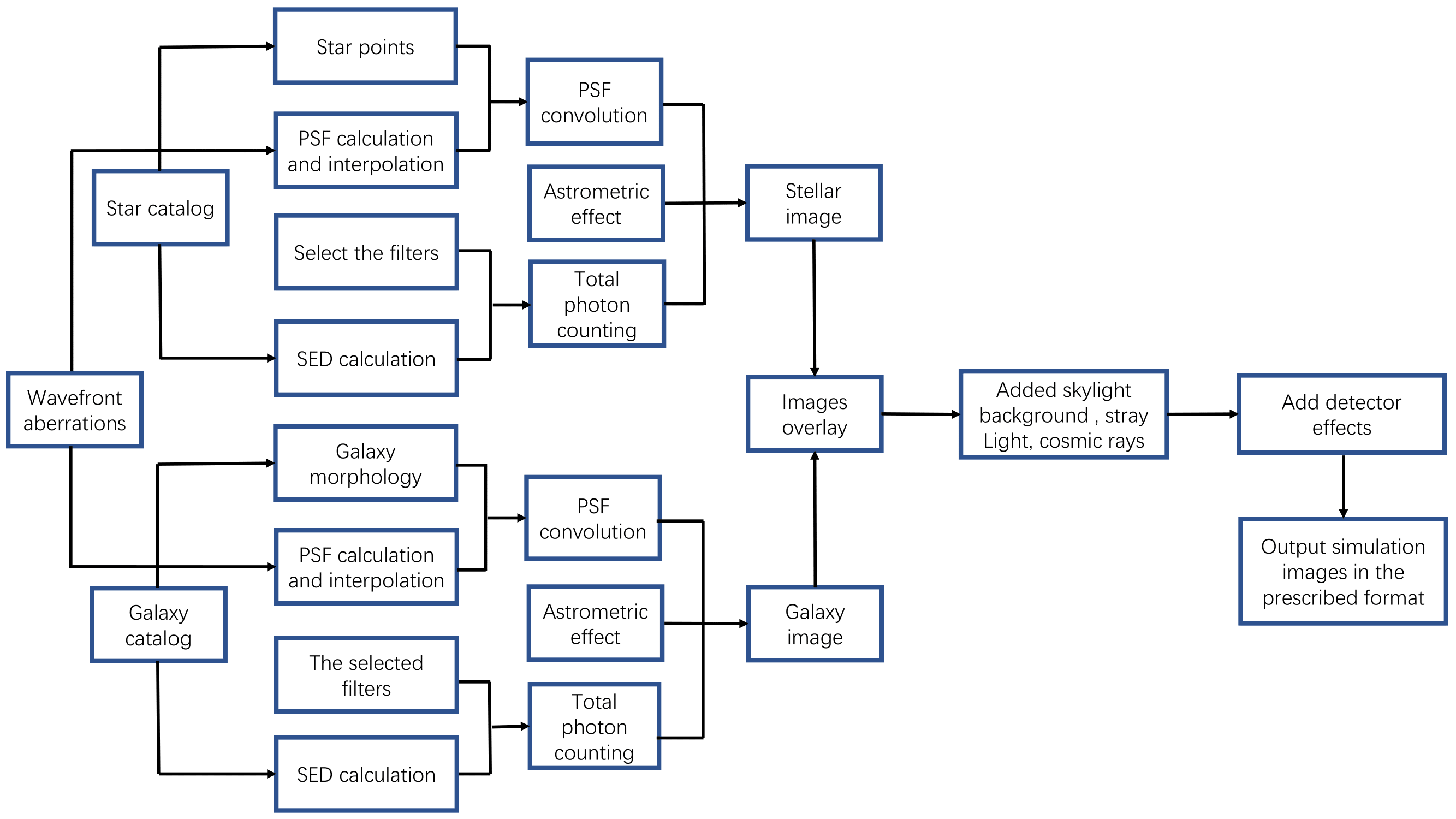}
   \caption{Schematic diagram of the simulation process of the MCI instrument}
   \label{Fig3}
\end{figure}
The flowchart of the MCI instrument simulation is shown in Figure~\ref{Fig3}. Stars are modeled as ideal point sources, and their two-dimensional intensity distributions on the CCD are primarily shaped by the corresponding PSFs. Their positions on the detector are determined mainly by celestial coordinates, the optical system, and astrometric effects. The total number of photons received from a star is governed by its spectral energy distribution (SED), the optical system's throughput, filter transmission, and the detector's quantum efficiency. In contrast, galaxies exhibit extended spatial profiles affected by both the PSF and gravitational lensing from foreground clusters. The Galaxy Morphology Simulation module accounts for these effects, including both weak and strong lensing. The photon count calculation for galaxies is identical to that for stars. The simulation software reads input catalogs and sequentially generates images of individual stars and galaxies, overlaying them on the detector frame to produce simulated images for the three channels. Additional components, such as the sky background, stray light, cosmic rays, and detector effects, are also incorporated. The final outputs are FITS files formatted according to the level 0 data definition, which serve as level 0 scientific image products.

\subsection{Wavefront aberrations of the MCI}

\begin{figure}
   \centering
 \includegraphics[width=\textwidth, angle=0, scale=0.99]{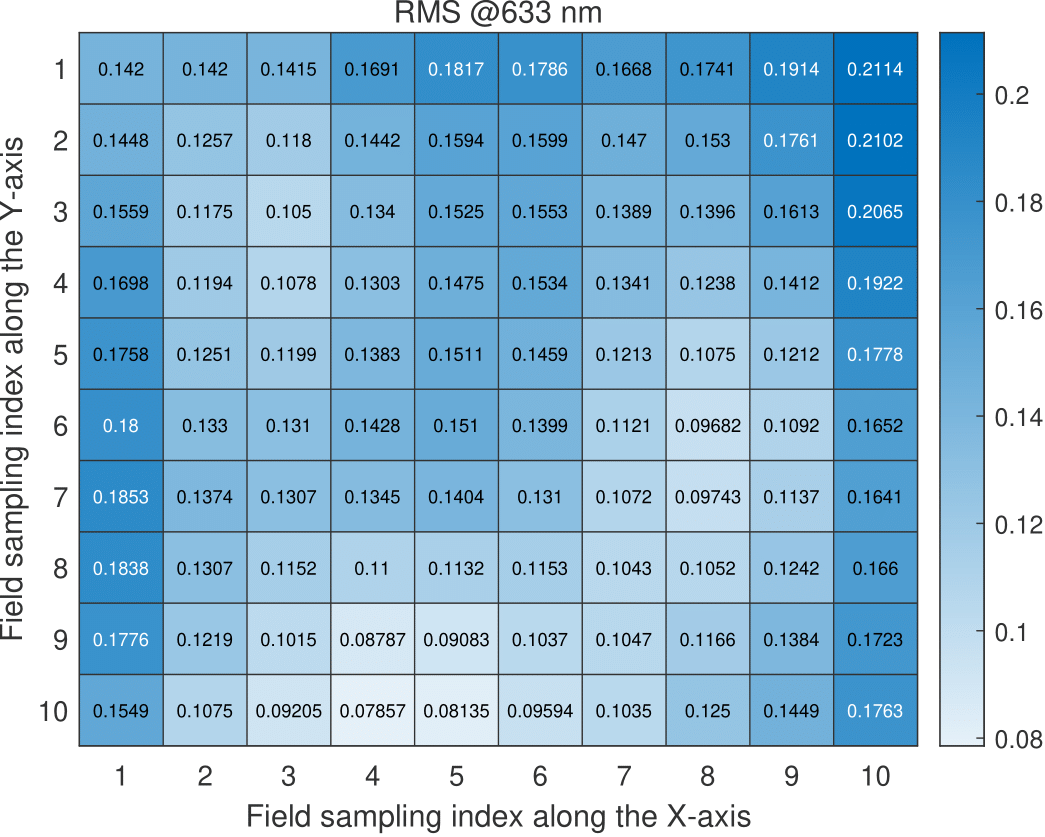}
   \caption{The simulated wavefront aberration distribution of the MCI at 633nm.}
   \label{Fig4}
\end{figure}
After considering various factors such as optical design errors, mirror processing errors, assembly errors, gravitational change errors and thermal deformation errors, we perform a simulation to obtain static aberration data covering the entire field of view of the MCI. A total of 100 sampled RMS values at the wavelength of \SI{633}{\nm} with a field of view sampling interval of 80\arcsec \ are shown in Figure~\ref{Fig4}. The RMS wavefront error does not reflect the image quality well, so we need to convert the wavefront error data into PSF data to evaluate the imaging quality of the system.

\subsection{PSF calculation}

Many science programs that involve MCI require an understanding of PSF. Examples include planet transit, planet detection, gravitational lensing studies, the host galaxy of the AGN, and high precision photometry in crowded fields. Therefore, understanding and constraining variations in the PSF core are crucial for predicting the scientific performance of the MCI. 
A PSF is a diffraction pattern caused by light from a point source passing through an optical system, with various optical components and obscurations adding structure to it. In the case of the MCI, there is no obscuration. Optical aberrations manifest themselves as errors in the optical path length. These properties can be simulated on a computer by creating a two-dimensional array containing the aperture function and another array containing the optical path difference (OPD) function. The OPD can usually be calculated as the sum of Zernike polynomials describing specific aberrations and additional terms such as local errors in the mirror surfaces. 
\begin{figure}
   \centering
 \includegraphics[width=\textwidth, angle=0, scale=0.99]{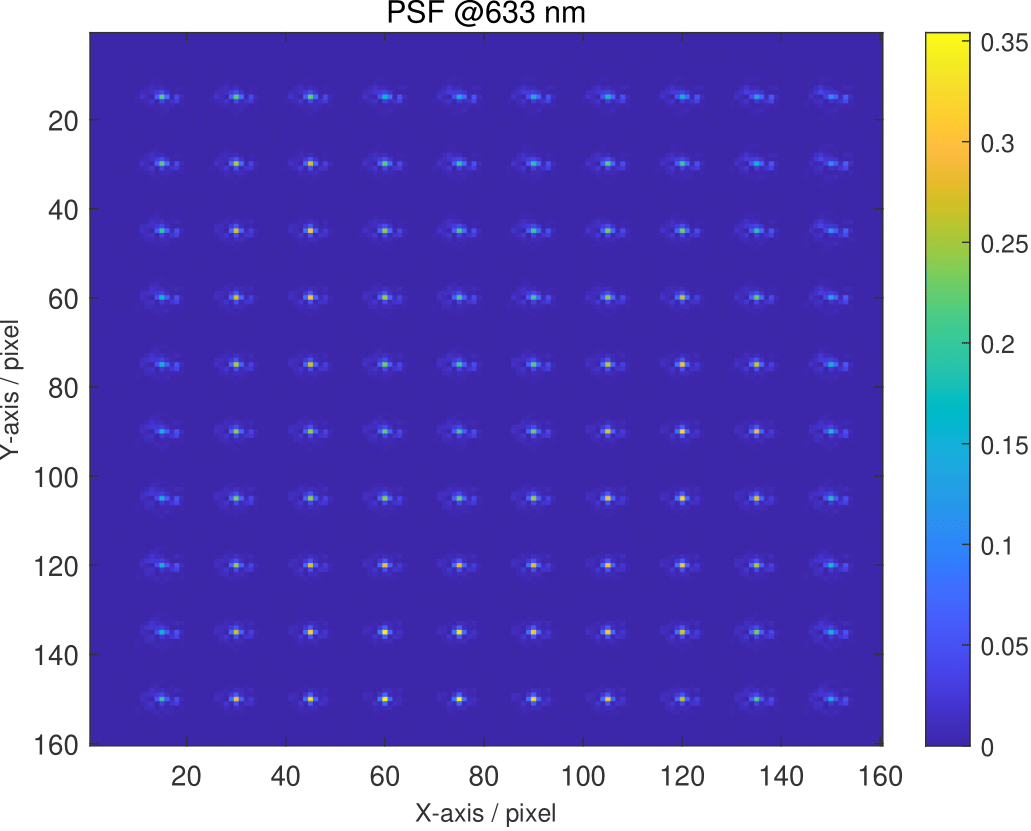}
   \caption{The simulated PSF data at the wavelength of 633nm.}
   \label{Fig5}
\end{figure}

These two functions are generated for each point on a grid of $N/2$ by $N/2$, providing critical sampling, where $N$ by $N$ are the dimensions of the arrays. The OPD and aperture function $A$ form the complex pupil function $P$:
\begin{center}
\begin{equation}\label{eq 1}
P=Ae^{(i \cdot 2\pi OPD/\lambda)}
\end{equation}
\end{center}
and its Fourier transform yields the amplitude spread function (ASF). Finally, squaring the modulus of the ASF gives the PSF:
\begin{equation}\label{eq 2}
PSF=|FFT(P)|^2,
\end{equation}
The Nyquist-sampled PSF for a specific wavelength (in which the OPD is defined) with the spacing between pixels being $F\lambda/2$, where $F$ is the focal ratio of the instrument, is then obtained. Thus, the sizes of the arrays used for generating the critical PSF depend on the required size of the integrated PSF and the wavelength. 
However, the FFT-based implementation has limitations for short wavelengths, requiring zero padding on the pupil plane and cropping operations on the focal plane. The uniform sampling interval in the two planes cannot accurately compute an undersampled PSF for short wavelengths. A more flexible method is to express the two-dimensional discrete Fourier transform (DFT) as a matrix triple product (MTP), which removes the constraints imposed by the traditional FFT. This allows for nonuniform sampling intervals in both the pupil plane and the focal plane. While a uniform sampling interval is often required on the pupil plane, the MTP operation provides flexibility for PSF calculations. Conventionally, the discrete computation is implemented via a 2D FFT algorithm. It can also be expressed in MTP form \citep{jurling2018}.

\begin{equation} \label{eq 3}
ASF=\Omega_yP\Omega_x,
\end{equation}
where $\Omega_x=e^{-i2\pi K_x^TX}$ and $\Omega_y=e^{-i2\pi K_y^TY}$. $K_x$, $K_y$, $X$ and $Y$ are the coordinates, represented by row vectors, in the spatial frequency domain and the spatial domain, respectively. $T$ represents the transpose. Eq.~\ref{eq 3} indicates that, in the MTP operation, the sampling intervals on both the pupil and focal planes can be irregular, despite the fact that a uniform sampling interval on the pupil plane is often required as a standard procedure for most PSF calculations. The normalized single wavelength (\SI{633}{\nm}) PSF simulation data are shown in Figure~\ref{Fig5}. An interval of 80\arcsec\ is used to cover the entire field of view of the MCI. The color bar shows the ratio of the energy of the central pixel to the total energy of the PSF. Notably, the pixel scale of this image is 0.05\arcsec\ . For visualization purposes, the PSF images from 100 field samples are compiled into a single composite image. Although the X-axis and Y-axis are labeled in pixel units for display convenience, their values should not be confused with the actual field sampling interval.

\begin{figure}
   \centering
 \includegraphics[width=\textwidth, angle=0, scale=0.99]{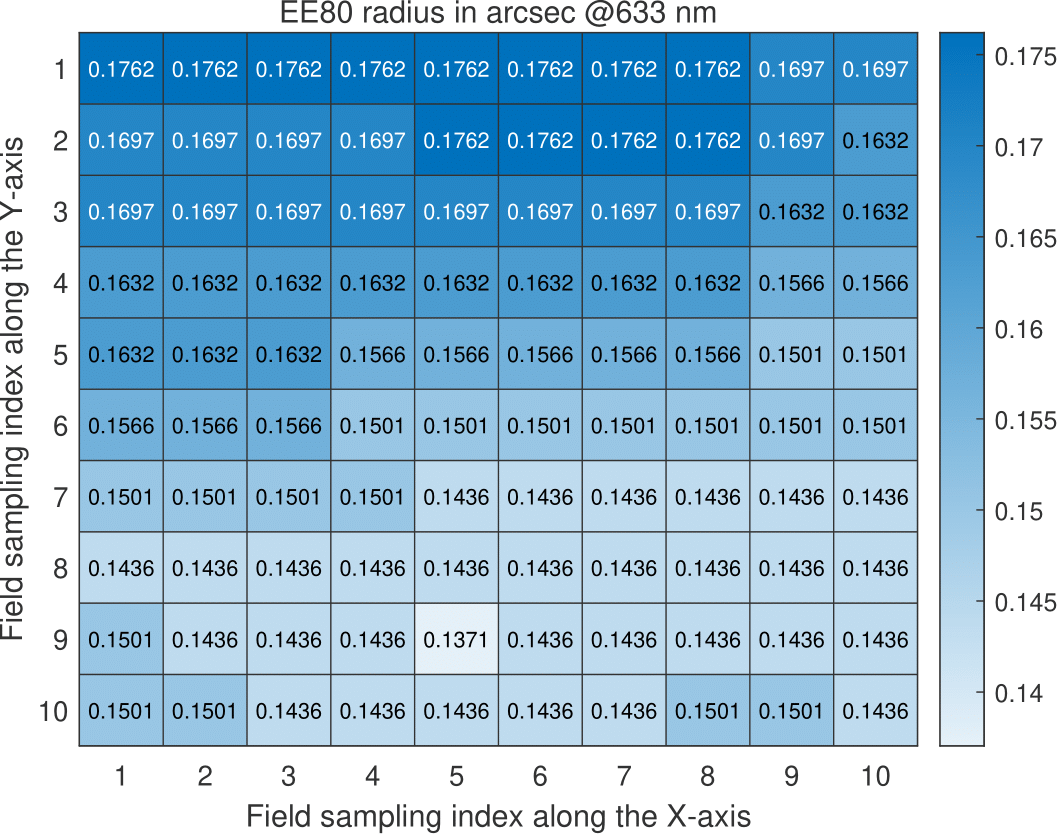}
   \caption{EE80 values of the simulated PSF data.}
   \label{Fig6}
\end{figure}

Experience with previous space telescopes—most notably the Hubble Space Telescope (HST)—has demonstrated that the potential for groundbreaking discoveries is closely tied to the quality and accurate characterization of the telescope’s PSF. During data analysis, a well-modeled and well-understood PSF is essential for extracting reliable scientific results. In this context, we present initial simulation results from our study of several simplified approaches to modeling the MCI’s PSF, with a particular focus on encircled energy. Encircled energy is defined as the fraction of total flux contained within a specified radius, with different radii serving different analytical needs. For the MCI, performance requirements specify the encircled energy within a radius of 180 milliarcseconds (mas). The simulated $80\%$ encircled energy (EE80) radii, shown in in Figure~\ref{Fig6}, meet the MCI’s optical design criteria at a wavelength of \SI{633}{\nm}. To ensure precise EE80 quantification, oversampled PSFs were employed for all encircled energy calculations. The resulting identical EE80 values at distinct field positions (Figure~\ref{Fig6}) thus represent intrinsic optical properties rather than computational artifacts. Notably, even aberrations with identical RMS wavefront errors can generate PSFs with divergent FWHM and EE80 characteristics, as these metrics depend not only on RMS magnitude but also on the components of the wavefront error and their coefficients. This explains why the field points with statistically significant differences in RMS wavefront error (Figure~\ref{Fig4}) can yield indistinguishable EE80 values (Figure~\ref{Fig6}).

\subsection{Filters and multi-wavelength PSF calculations}

\begin{figure}
   \centering
 \includegraphics[width=\textwidth, angle=0, scale=0.99]{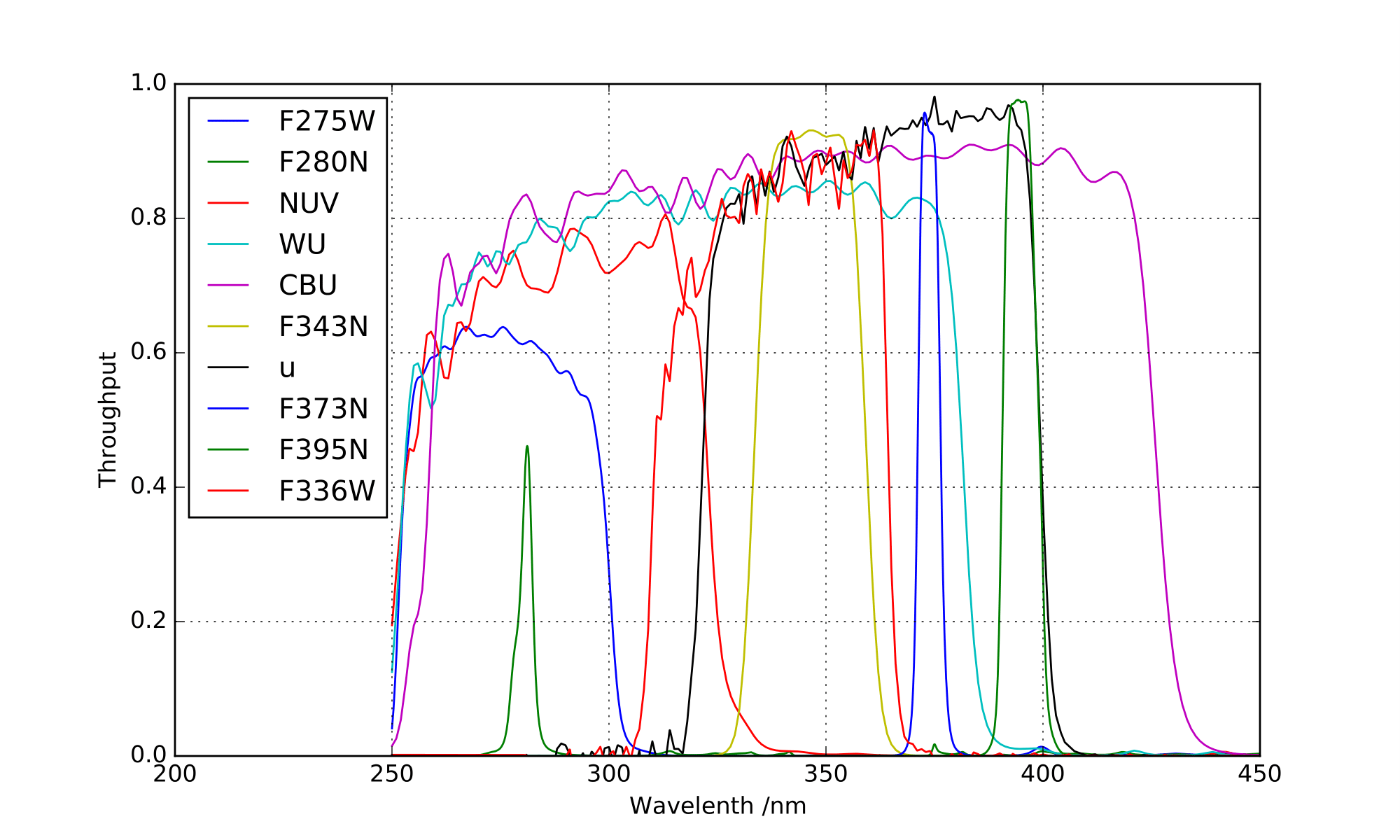}
   \caption{Throughput for the filters in the ultraviolet channel.}
   \label{Fig7}
\end{figure}

\begin{figure}
   \centering
 \includegraphics[width=\textwidth, angle=0, scale=0.99]{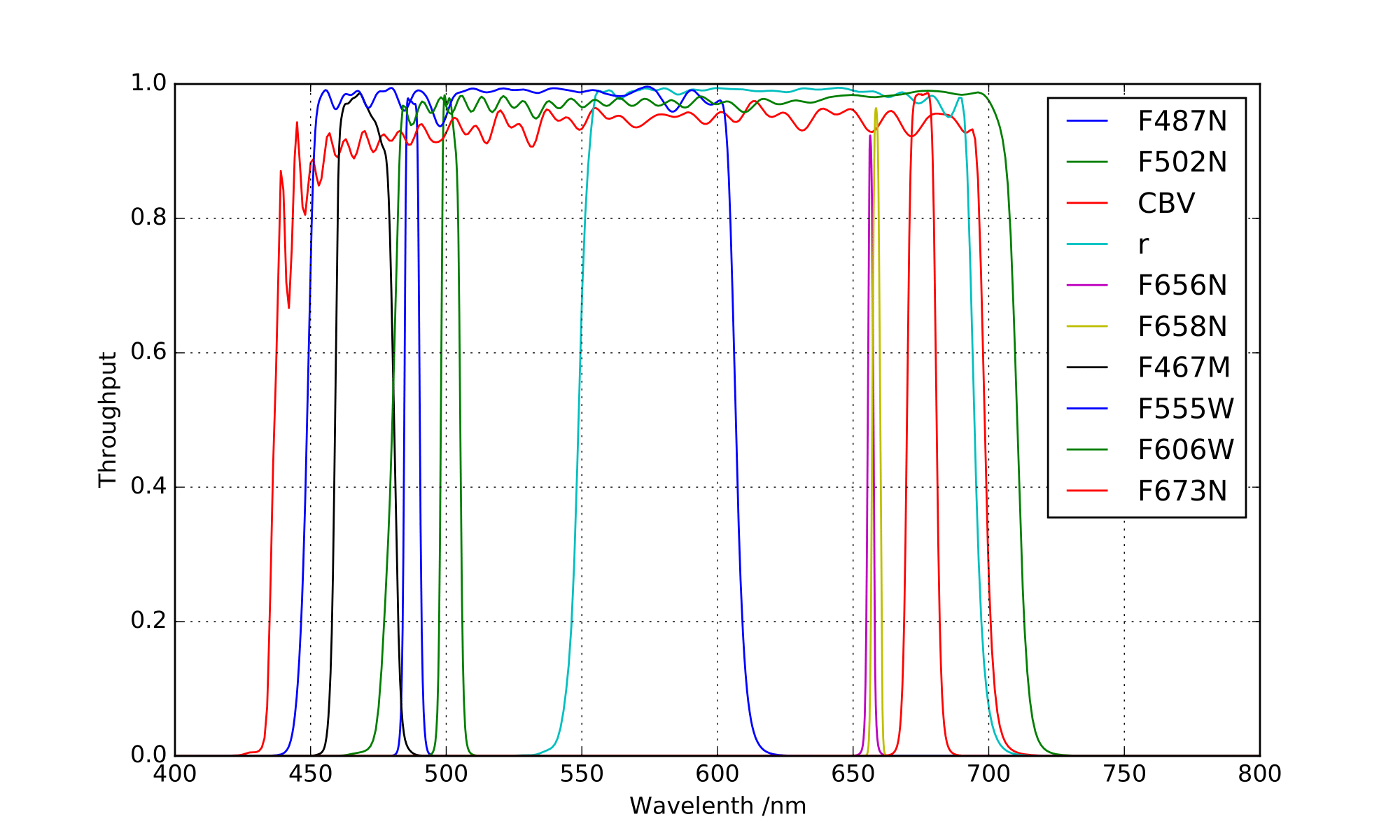}
   \caption{Throughput for the filters in the visible channel.}
   \label{Fig8}
\end{figure}

\begin{figure}
   \centering
 \includegraphics[width=\textwidth, angle=0, scale=0.99]{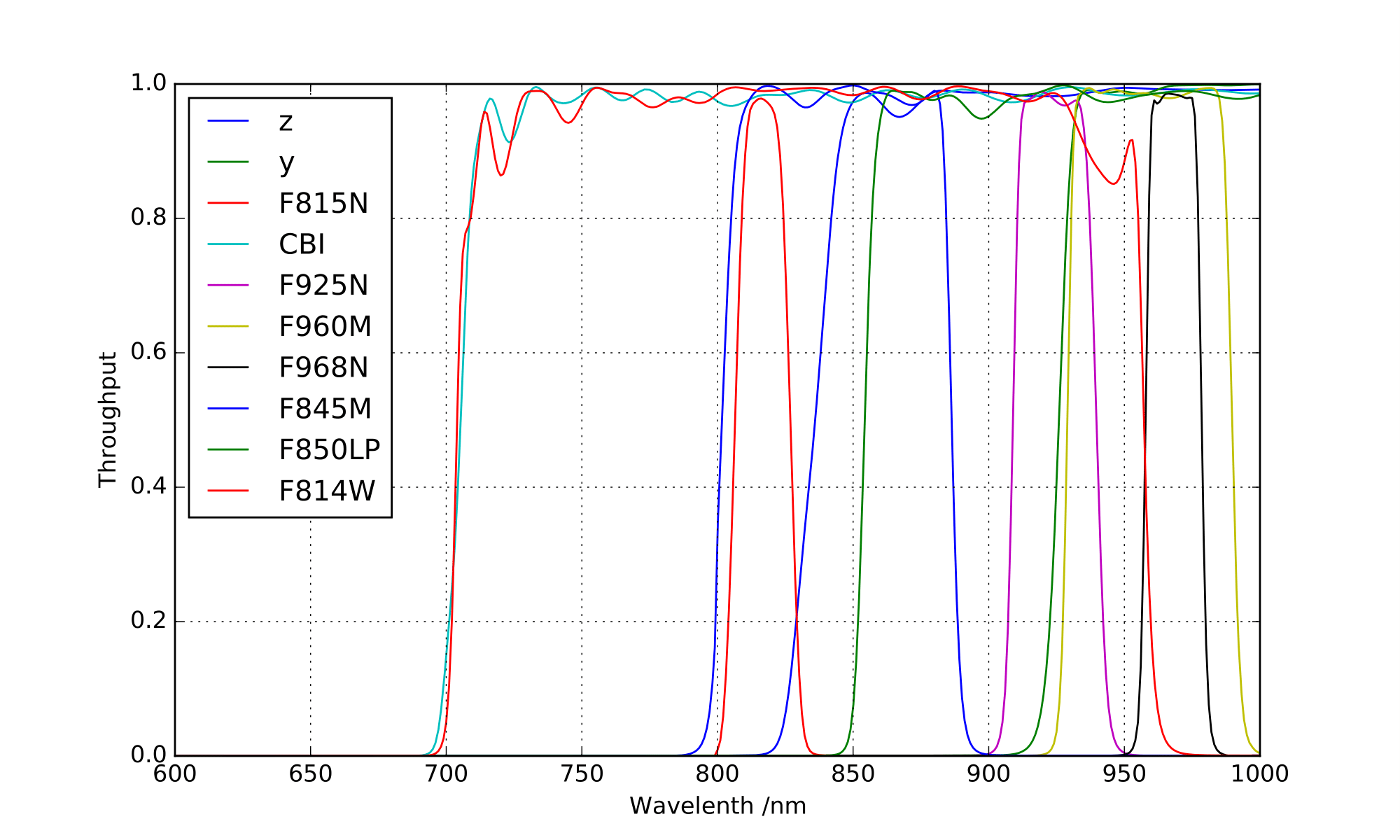}
   \caption{Throughput for the filters in the near-infrared  channel.}
   \label{Fig9}
\end{figure}
As the PSF varies with wavelength, it is essential to generate a number of PSFs via the above technique at wavelengths that sample the filter to account for the filter's bandpass. The MCI has three optical channels, including an ultraviolet channel, a visible channel and a near-infrared channel. Each channel has ten filters for different scientific objectives, and the measured throughput curves for engineering samples of the filters are shown in Figure~\ref{Fig7} to Figure~\ref{Fig9}.
The above calculation is specified for a single wavelength, but broadband PSFs can be derived as weighted sums of monochromatic PSFs within the bandpass. The weights applied to each wavelength's PSF are calculated by combining the source spectrum (for simulated targets), filter throughput, system optical efficiency, and detector efficiency. The simulated image of a source (star or galaxy) can then be calculated as
\begin{equation}\label{eq 4}
IMG=S_{Phon}*I*(PSF_1*C_1+PSF_2*C_2+...+PSF_W*C_W),
\end{equation}
where $C_1, C_2, ..., C_W$ are the energy normalization coefficients, which are related to these factors including the flux of the source, the optical efficiency of the optical system, the filter efficiency and the detector efficiency. $S_{Phon}$ represents the total number of photons determined by the flux of the input source as well as the optical efficiency of the optical system, the filter, the detector efficiency, and the exposure time. $I$ is the normalized two-dimensional intensity distribution of the input source, and is the delta function if the input source is a star. $W$ is the number of the sampled wavelengths of the filter. In the simulation, each filter has 7 sampled wavelengths, and their values are defined as
\begin{equation}\label{eq 5}
\lambda_n=\lambda_0+(n-4)\frac{\lambda_{FWHM}}{4}, n=1,...7,
\end{equation}
where $\lambda_0$ is the central wavelength of the filter and $\lambda_{FWHM}$ is the FWHM of the bandpass of the filter.

\subsection{PSF interpolation}

For accurate quality characterization of simulated scientific images, the PSF at the target source location must first be generated by interpolating the laboratory PSF field data. This interpolated PSF is then convolved with the target source distribution to simulate the multicolor photon distribution. In the current version of image simulation, the classical inverse distance weighting (IDW) interpolation method is used preferentially to obtain an accurate PSF at the target source location. IDW is one of the oldest yet most widely used spatial interpolation methods \citep{Gentile_2012,Shepard1968ATI}. The fundamental concept behind the IDW interpolation technique is that in a field with continuous variation in PSF, the PSF at any given position can be interpolated by the adjacent PSF samples, and the corresponding interpolation weight is inversely proportional to the distance between the target source and the samples. The estimated value PSF($x_0$) at a target point $x_0$ is given by 
\begin{equation}\label{eq 6}
PSF(x_0)=\sum^{M}_{k=1}w_k PSF(x_k),
\end{equation}
where $w_k=\frac{1}{r^{\beta}(x_0, x_k)}/\sum^{M}_{k=1}\frac{1}{r^{\beta}(x_0, x_k)}$, $\beta$ is a power parameter and $r(x_0, x_k)$ represents the distance between the target source $x_0$ and the $k$-th adjacent PSF sample. $M$ denotes the number of points located in a particular vicinity surrounding the designated point. The power parameters ranging from 1 to 4 are commonly selected, with the most chosen option being 2, which results in the inverse distance-squared interpolator. To evaluate the influence of the IDW algorithm on the accuracy of image simulation, a one-to-one cross-test was carried out on the PSFs before and after interpolation. The results indicate that the relative error of the PSF at the interpolation point in the test region is less than 1\%.

\subsection{Geometric distortion}
\begin{figure}
   \centering
 \includegraphics[width=\textwidth, angle=0, scale=0.99]{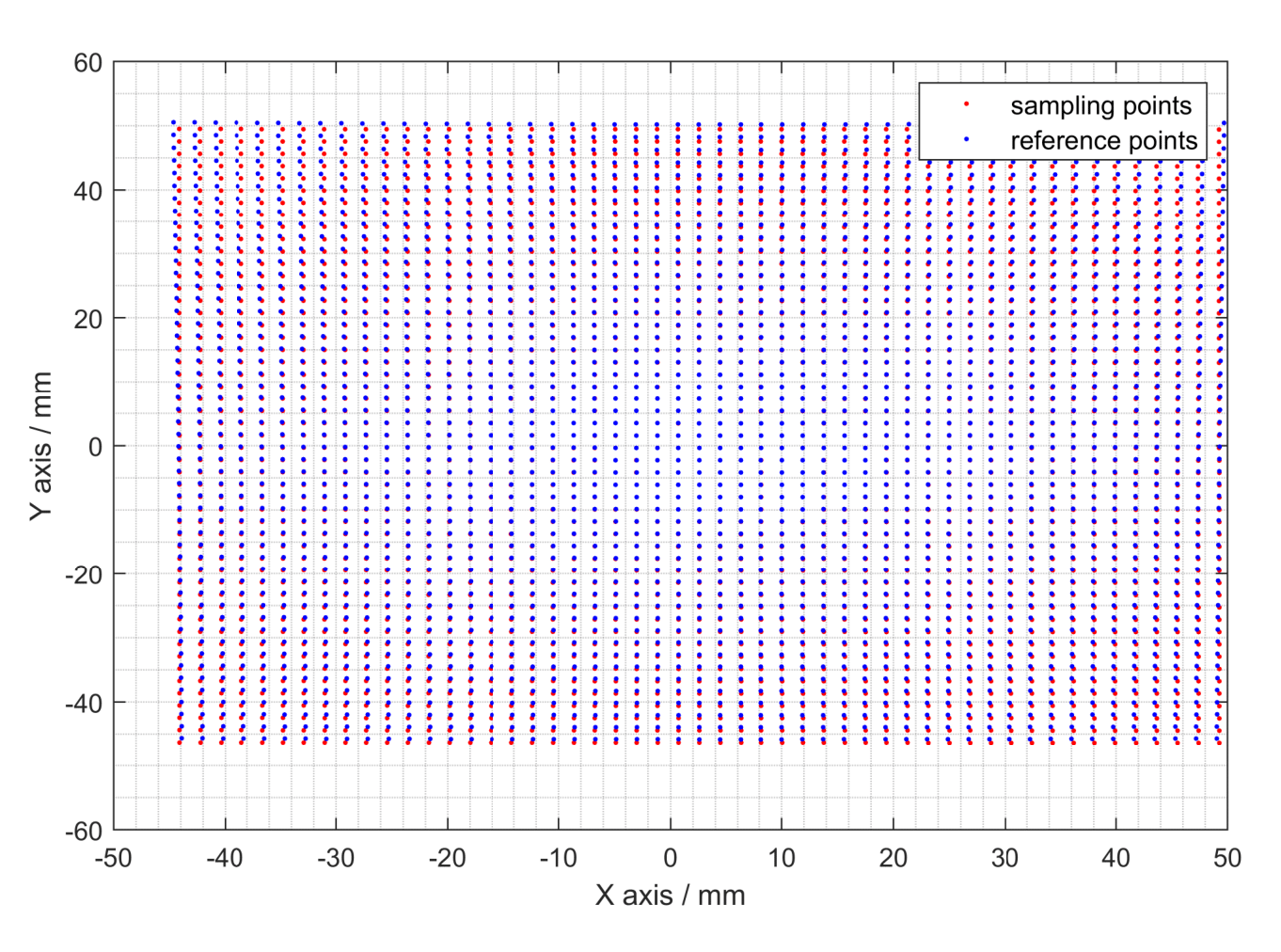}
   \caption{Distortion sampling points (blue dots) and reference sampling points (red dots).}
   \label{Fig10}
\end{figure}
The geometric distortion model is also based on the simulated PSF data. In the absence of any geometric distortion, the celestial coordinates are mapped to the UV plane by WCS projection, and the corresponding pixel coordinates $(x_0, y_0)$ of the object are then obtained. In the case of geometric distortion, the mapping relationship between the UV plane and the pixel coordinates is altered, resulting in the corresponding pixel coordinates of the object being obtained. This is achieved by establishing a mapping relationship between $(x_0, y_0)$ and $(x, y)$. The geometric distortion data are exported through Python joint programming with Zemax, and the geometric distortion data in the visible channel are shown in Figure~\ref{Fig10}. The full field of view comprises $50 \times 50$ sampling points, with red and blue points marking reference and distorted positions, respectively. The geometric distortion is modeled using fifth-order polynomials to establish the mapping relationship between $(x_0, y_0)$ and  $(x, y)$, thereby simulating the image field distortion effect. By applying the same modeling approach, the geometric distortion mapping functions for the other two channels can be derived.

\subsection{Star image simulation}

\begin{figure}
   \centering
 \includegraphics[width=\textwidth, angle=0, scale=0.99]{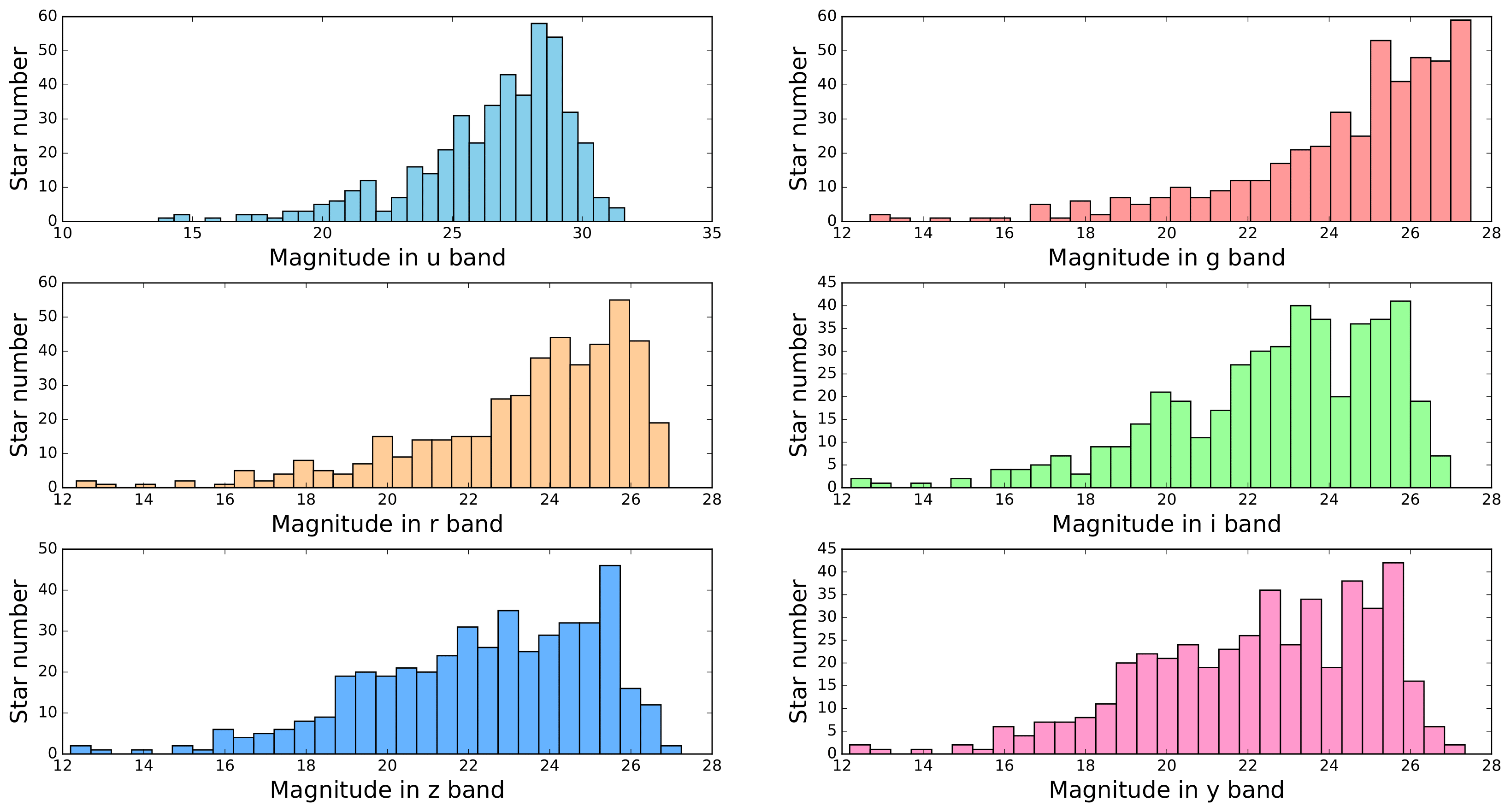}
   \caption{Star magnitude distributions in different bands.}
   \label{Fig11}
\end{figure}

The currently entered star catalog consists of simulated and real data from Gaia. The bright-end stars (\textless 22 mag) in the input synthetic catalog are derived from the Gaia DR3 catalog\citep{2023A&Agaia}. The faint-end stars (\textgreater 22 mag) are derived from the CSST-TRILEGAL catalog\citep{Chen_2023}. The typical single exposure time of the MCI is designed to be 300 seconds. The telescope points to the field with $Ra=116.1808^\circ$ and $Dec=39.4231^\circ$. Figure~\ref{Fig11} illustrates the distribution of star magnitudes across various spectral bands within the MCI's field of view. It meets the accuracy requirements for MCI deep field detection for faint-end stars up to 32 mag. Moreover, the Gaia dense field catalogs will be utilized to generate simulated images of dense star fields for the purpose of evaluating astrometric algorithms. Furthermore, the testing of flux calibration algorithms will be facilitated by the utilization of catalogs of white dwarfs containing SEDs.

The SED of a single star is also simulated, which is generated with the Gehong module, as discussed in a companion paper (Feng et al., in preparation 2025). The input parameters of the Gehong module include the distributions of two-dimensional physical parameters of galaxies; physical parameters of the stellar continuum (magnitude, mean age, metallicity, velocity, velocity dispersion, dust extinction); and physical parameters of ionized gas emission lines (Halpha flow, metallicity\citep{Sanchez2014}, velocity, velocity dispersion, dust extinction, and kinematics\citep{Cappellari2006,Genzel2011}. 

\subsection{Image simulation of galaxies}
\label{Image simulation of galaxies}
We employ a large-scale dark matter-only simulation with 4.1 trillion particles to construct the mass distribution of cosmic structures and mock galaxy catalogs for deep field surveys including blank deep fields and cluster deep fields (Wang et al. in preparation)

The simulation spans 2.5 $h^{-1}$ Gpc on each side, with an initial redshift of $z=127$, and a mass resolution of $3.216 \times 10^{8} h^{-1} \rm M_{\odot}$. This resolution is comparable to current large-scale simulations used in surveys such as DESI and Euclid, including the massive simulations in the Uchuu and Millennium-TNG projects\citep{2021MNRAS.506.4210I, 2023MNRAS.524.2556H, 2024arXiv240513495E}. To capture the details of the halo profiles and substructures accurately, we set the spatial resolution (softening scale) to $3.0 h^{-1} \rm kpc$.

The simulation is performed via the latest version of the $N$-body code, PhotoNs-3.7, which incorporates a Particle-Mesh Fast-Multipole-Method gravity solver and heterogeneous acceleration\citep{Wang_2021, Wang_2021_gpu}. This version improves memory management for parallel computations and optimizes both the data structure and particle memory usage. The initial conditions are generated using a Zel'dovich shift algorithm within PhotoNs-3, ensuring high resolution from the start at $z=127$\citep{2022MNRAS.517.6004W}. The initial power spectrum is derived from CAMB in a flat $\Lambda$CDM cosmology model with $\Omega_m=0.3111$, $\Omega_{\Lambda}=0.6889$, and $H_0 = 67.66 \rm km/s, Mpc$. The power spectrum index is 0.9665, and $\sigma_8 = 0.8102$\citep{2020A&A...641A...6P}.

\begin{figure}
   \centering
 \includegraphics[width=\textwidth, angle=0, scale=0.7]{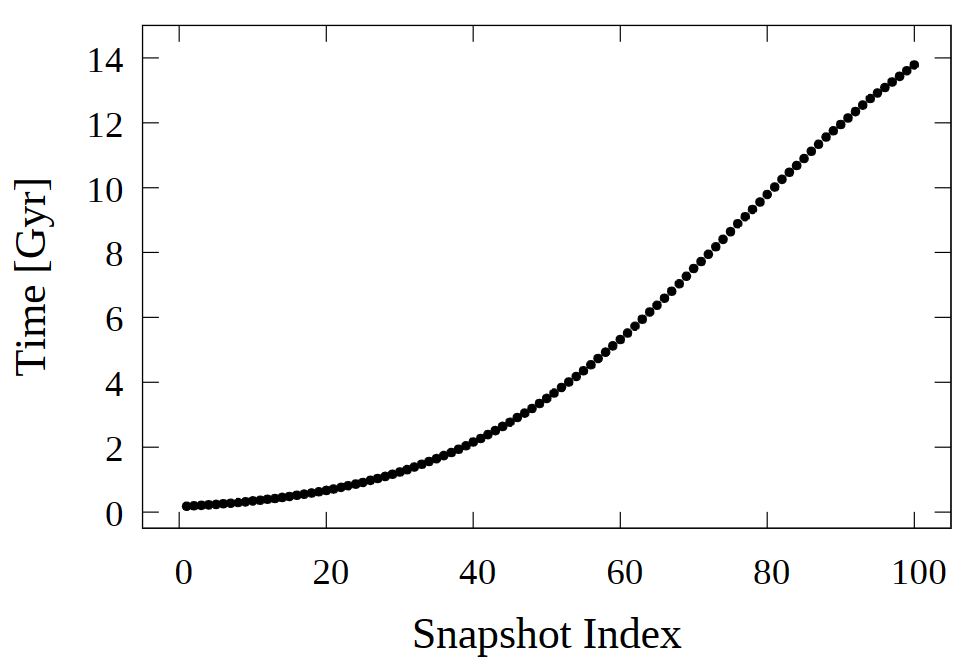}
   \caption{Output timeline of the N-body simulation}
   \label{NbodySim}
\end{figure}

A total of 100 full snapshots are stored from $z=30$ to $z=0$, totaling over 12 petabytes of data, with 32-bit positions, 32-bit velocities, and long int IDs, as shown in Figure~\ref{NbodySim}. The mass distribution and mock galaxy catalog are constructed from slices of these simulation particles.

The blank field mock will utilize the aforementioned simulation setup. The cluster deep field mock further includes intra-cluster light (ICL) and gravitational lensing effects and also simulates the SEDs of galaxies, as detailed in the companion paper by Xie et al. (in preparation 2025).

\subsection{Other instrumental effects}

\subsubsection{Comprehensive efficiency}
\begin{figure}
   \centering
 \includegraphics[width=\textwidth, angle=0, scale=0.99]{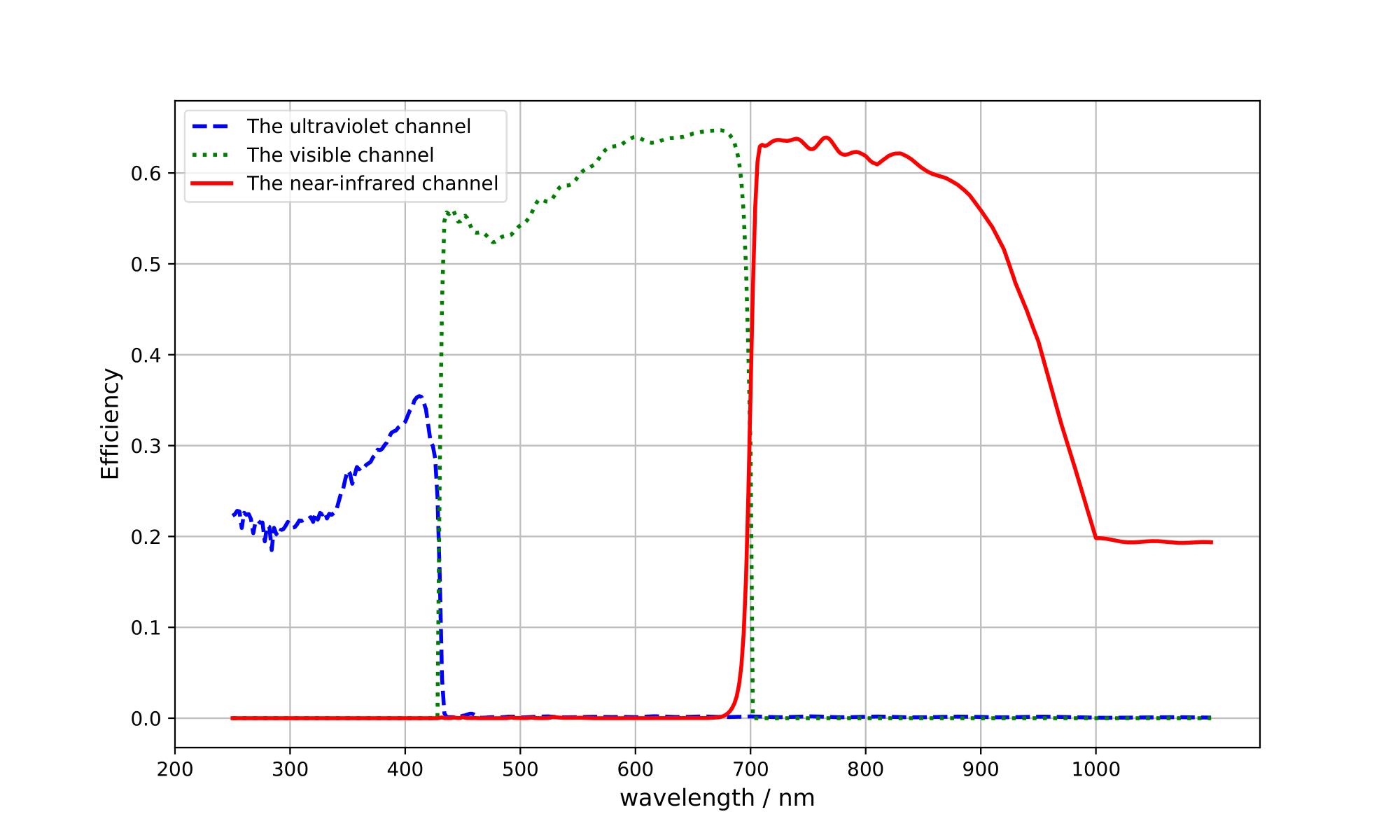}
   \caption{ Efficiency curves for three channels of the MCI.}
   \label{Fig13}
\end{figure}
Figure~\ref{Fig13} shows the efficiency curves in the three channels used by the simulation program to calculate the number of photons observed at the target source. These efficiency curves consider the optical efficiency of the entire optical system, the transmittance of the dichroic mirrors, and the quantum efficiency of the CCD detectors.

\subsubsection{Shutter effect}
\begin{figure}
   \centering
 \includegraphics[width=\textwidth, angle=0, scale=0.99]{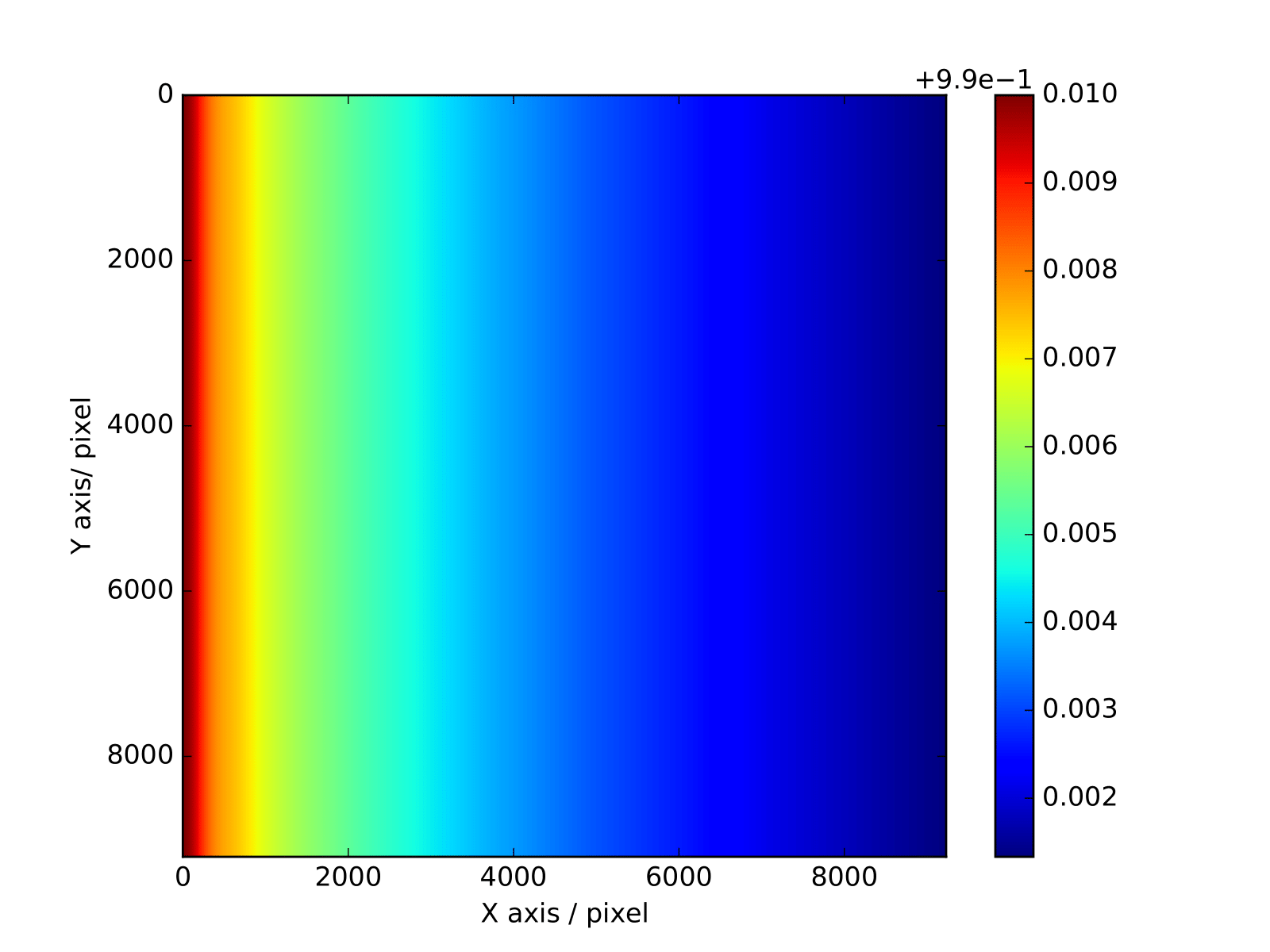}
   \caption{Shutter effect matrix map.}
   \label{Fig14}
\end{figure}
The simulation also considers the influence of the shutter effect caused by the opening and closing of the shutter. The exposure times vary at different positions on the focal plane because of this effect. The simulated shutter opening and closing times are both 1.3 seconds. As the angular acceleration changes with time following a positive cosine curve during the shutter operation, the shutter effect is the same during opening and closing. The shutter effect for a single frame exposure can be 2.6 seconds. Currently, the shutter effect is only simulated for exposures longer than 2.6 seconds and is not simulated in the dark and bias images. Assuming that the exposure time is the same at all points on the focal plane parallel to the shutter bearing, the shutter effect remains constant in the direction parallel to the bearing. However, the exposure time differs at locations perpendicular to the bearing, resulting in uneven exposure. The shutter effect matrix simulating a 300 second exposure is shown in Figure~\ref{Fig14}.

\subsubsection{Cosmic rays}

The simulation of cosmic rays is based on cosmic ray data recorded from real observations by the Hubble Space Telescope. These data include information on the length and energy of the cosmic rays. The total number of cosmic rays in each detector pixel is used as a configurable input parameter for the simulation. In the simulation, the cosmic ray coverage is set to $0.1\%$ for an exposure time of 300 seconds. The position at which cosmic ray particles hit the detector follows a uniform random distribution, and the angle of bombardment is also random. When a particle hits the detector, it leaves a cosmic ray trace, and the energy of the cosmic ray decays according to a power-law intensity.

\subsubsection{Photo Response Non-Uniformity}

Photo-response non-uniformity (PRNU) represents the uniformity of a camera’s response to light, which is particularly important in high-light applications. It is defined as the standard deviation of the gain values of the pixels and is expressed as a percentage. The simulation uses a two-dimensional Gaussian distribution to simulate the non-uniformity of the normalized response between pixels, with a mean of 1 and a standard deviation of $0.1\%$.

\subsubsection{Brighter-fatter and charge diffusion}

Bright spots in the image appear slightly wider than faint ones because of the brighter-fatter effect of the detector. In addition, all CCDs suffer from charge diffusion, resulting in image blur. CCD pixels are not physically separate elements; they are defined by the electromagnetic fields generated by the electrode structure on the detector surface. Pixel edges are therefore inherently fuzzy. When a photon is converted to an electron in the substrate, the electron tends to be attracted to the nearest electrode. However, in regions where the field strength is weak, the electron can travel some distance from its point of generation, possibly even into an adjacent pixel, causing charge diffusion. We use Galsim's built-in silicon sensor module to simulate brighter-fatter and charge diffusion effects. The underlying implementation principle uses the Poisson equation solver and the Monte Carlo method to simulate the drift process of photoelectrons in the electric field \citep{2021Craig}. For this simulation, we used the measured parameters of the ITL STA3800C CCD detector selected by the LSST (Large Synoptic Survey Telescope) as input parameters to the silicon sensor module.

\subsubsection{Astrometric effect}
To accurately simulate the apparent line-of-sight directions of stars as observed by the CSST during its orbital operations, the positions of stellar objects are systematically transformed using the microarcsecond-precision relativistic astrometric model developed by \citealt{Klioner2003AJ}. This procedure converts stellar coordinates from the International Celestial Reference System (ICRS) at a reference epoch to their apparent directions in the Geocentric Celestial Reference System (GCRS) at the observation epoch. The main steps are as follows:

\begin{enumerate}
    \item \textbf{Proper Motion Correction:} Based on stellar proper motions and, when available, radial velocities, the positions are propagated from the catalog’s reference epoch (e.g., J2000 ) to the epoch of observation, yielding the stars’ instantaneous spatial positions with respect to the Barycentric Celestial Reference System (BCRS).

    \item \textbf{Parallax Correction:} The annual parallax effect is applied together with the precise BCRS position vector of the observer at the observation epoch, which is obtained from high-accuracy planetary ephemerides. This transforms the barycentric direction into a geometric direction from the geocenter while retaining its expression in BCRS coordinates.

    \item \textbf{Gravitational Light Deflection:} The observed propagation direction of the starlight is adjusted to account for relativistic deflection due to gravitational fields—primarily the dominant deflection caused by the Sun and, to a lesser extent, by other massive bodies in the solar system.

    \item \textbf{Aberration:} The apparent directional shift caused by the CSST’s motion with respect to the solar system barycenter is corrected, yielding the final observed position in the GCRS at the observation epoch.
\end{enumerate}

This transformation workflow conforms to the IAU (2000) resolutions on astrometric reference systems and achieves microarcsecond-level accuracy while remaining computationally practical. For a full theoretical background, see \citealt{Klioner2003AJ}.

\iffalse
The astrometry module used in the simulation can convert the position of the input catalog (which, by default, has astrometric parameters under the ICRS system) to the real observation position of the telescope. The astrometry module follows the convention defined by the International Astronomical Union (IAU) in its official specification (IERS2021). The accuracy of the astrometric model used is up to 10 mas. In the calculation process, the astrometry module requires knowledge of the satellite's orbital parameters (three-dimensional spatial position parameters and three-dimensional velocity parameters) and the star's motion information (RAJ2000, DecJ2000, proper motion, radial velocity, parallax) at the time of the observation. After passing through the astrometry module, the target position can be changed by more than 20\arcsec. When dealing with galaxy positions, proper motion, radial velocity and parallax are all set to 0.
\fi

\subsubsection{Background and stray light}

The amount of background emission is critical to the depth of many MCI images. There are three main components of the background, namely, zodiacal light, earth shine and scattered light. The magnitude of the zodiacal light is influenced by the angle of the target relative to the Sun and the ecliptic. Data from Leinert \citep{leinert19981997} are used to simulate zodiacal light. Earth shine varies greatly as a function of the angle between the target and the Earth's limb. The data from Figure 9.1 in the HST/WFC3 instrument manual (\url{https://hst-docs.stsci.edu/wfc3ihb/chapter-9-wfc3-exposure-time-calculation/9-7-sky-background, https://cads.iiap.res.in/tools/zodiacalCalc/Documentation}) are used in our earth-shine simulations. Stray light is generally used to describe unwanted light that enters the focal plane of an optical system and is a critical factor that has a significant impact on the high precision photometry of optical astronomical telescopes, as it has the potential to reduce the signal-to-noise ratio of astronomical objects. Stray light analysis and simulation are performed by the CSST Scientific Data Systems team, starting with the 3D modeling of the optical and mechanical geometry of the CSST and the assignment of surface properties via ray tracing methodology.  

\subsubsection{Charge transfer inefficiency}

The high flux of damaging radiation at the orbital altitude of space telescopes creates an ever-increasing population of charge traps in the silicon of charge-coupled devices (CCDs), reducing their charge transfer efficiency (CTE), which is quantified by the fraction of charge successfully transferred (clocked) between adjacent pixels. The term charge transfer inefficiency (CTI) is actually more useful in many cases. The main observable consequence of CTI is that a star whose induced charge must travel across several pixels before being read out will appear fainter than the identical star visible close to the readout amplifier. This effect is significant for all the CCD detectors used on the CSST instruments. The main causes of CTI trailing are the presence of electronic traps in the pixel and the electronics in the readout process. For the CTI effect added to the MCI, see the trailing profile of the CTI effect measured in the reference paper \citep{anderson2010empirical}. In the current simulation, the CTI trailing direction is set along the column direction, and the trailing direction can also be set along the row or column direction in the module. At present, the trailing length is generally 10 pixels, and the energy ratio of the trailing is approximately $0.003\%$ (the detector model used by the MCI is E2V CCD 290-99, and its simulated CTE is approximately $99.997\%$). Notably, the $99.997\%$ CTE value is a conservative estimate for algorithm validation, not based on radiation tests. This value was intentionally chosen to introduce a noticeable pixel trailing effect, which is beneficial for the development and validation of CTI correction algorithms. We have an internal research team dedicated to CTI studies, encompassing both simulation and correction methodologies. In parallel, this team is collaborating with domestic detector development groups to carry out radiation testing. The results of these ongoing efforts will be presented in future publications.

\subsubsection{Nonlinearity}

A CCD consists of an array of pixels on a very thin layer of silicon. Each incoming photon is converted into an electron-hole pair. Up to 50000-500000 electrons can be stored per pixel, depending on the CCD. After exposure, a controller can move the electrons around and read them out, converting the electron charge of each pixel into digital counts (analog-to-digital units or ADUs). The term "gain" refers to the conversion factor (electrons/ADU). Contrary to popular belief, CCDs are not perfectly linear systems. The number of ADU counts is not exactly proportional to the number of incident photons after bias subtraction. The nonlinearity of the controller gain is most likely due to its non-constant relationship with the number of electrons. There will be much greater nonlinearity as the CCD approaches saturation because there is a lower chance of capturing photons in near full pixels. The nonlinearity of the detector is simulated via the following formula:
\begin{equation}
f(x)=x-\alpha x^2,
\end{equation}
where $x$ represents the counts received by the pixel, $\alpha$ is the nonlinearity coefficient with the default value of $10^{-7}$ and $f(x)$ represents the simulated counts for this pixel affected by the nonlinearity effect of the detector.

\subsubsection{Other detector effects}

Each MCI sensor has a pixel size of $10 \mu m$ and an image area of $9216 \times 9232$ pixels. The sensors have registers at both the top and bottom, and each sensor has eight outputs for short readout times. Therefore, each chip has sixteen readout channels with different simulated gains, dark current noise, readout noise and bias values. Different readout channels have different gain values, and the typical value is 1.5. The typical value of the dark current noise in each output channel is $0.001~\rm e^{-1}/sec/pix$. The typical readout noise with a Gaussian distribution varies from $4.2$ to $4.7~\rm e^{-1}/sec/pix$ for different channels. The typical bias values vary from 400 to 640 for different channels. The camera is equipped with a 16-bit resolution, and the maximum gray value is 65535. If the maximum grayscale value of a pixel exceeds the maximum value of the pixel, the electrons present in that pixel become saturated and spill over to both sides of the pixel. The simulation takes into account bad and hot pixels. The final FITS files containing the three-channel image outputs are simulated, and the data structure is mapped to the level 0 CSST scientific data format.

The majority of instrument effects or function switches in the simulation can be set by the user. A comprehensive list of user-adjustable parameters and their functional descriptions can be found in the reference appendix for configurable instrument controls.

\section{MCI simulation results}
\label{sect:5}

Simulations via the MCI instrument software demonstrate its ability to generate calibration images, including dark-field, flat-field, and bias frames. The software also simulates observation images for various astronomical sources. This section presents simulated star and galaxy catalog images generated with the CSST-MCI simulation software. Additionally, we present the preliminary results from processing the software-produced CSST-MCI Class 0 science data via the MCI’s two core data processing pipelines: the flux calibration pipeline and the image coaddition pipeline.

\subsection{Star image simulation results}
\begin{figure}[htpb]
   \centering
 \includegraphics[width=\textwidth, angle=0, scale=1.0]{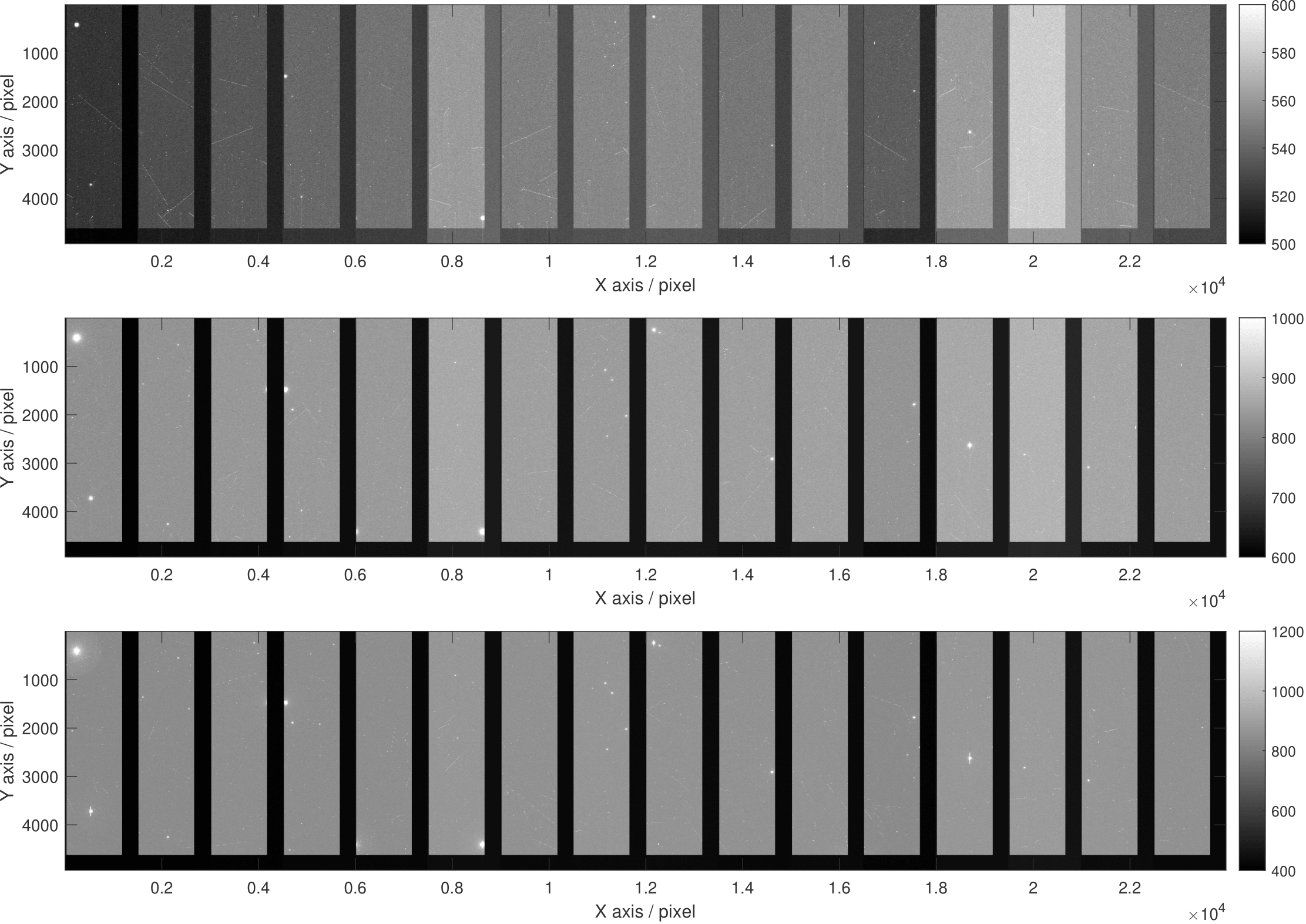}
   \caption{A demonstration of level 0 scientific data images obtained from the MCI's three-channel star image simulation. The ultraviolet channel (top). The visible channel (middle). The near-infrared channel (bottom).}
   \label{star_img}
\end{figure}

Figure~\ref{star_img} demonstrates level 0 scientific data images from the MCI's three-channel star image simulation. This simulation uses a combined star catalog incorporating both simulated data and real Gaia data. Bright stars (\textless 22 mag) originate from the Gaia DR3 catalog \citep{2023A&Agaia}, whereas faint stars (\textgreater 22 mag) derive from the CSST-TRILEGAL catalog \citep{Chen_2023}. The three channels employ 'CBU', 'CBI', and 'CBV' filters, respectively, with a single exposure time of 300 seconds targeting the field at $Ra=116.1808^\circ$ and $Dec=39.4231^\circ$. The simulated images clearly reveal saturation spillover from bright stars alongside cosmic rays captured by the detectors. Given that each channel detector utilizes 16 readout circuits, each simulated image comprises 16 distinct blocks. Initial processing can restore these segmented images into the original composite frame for subsequent data analysis.

\subsection{Galaxy image simulation results}
\begin{figure}[htpb]
   \centering
 \includegraphics[width=\textwidth, angle=0, scale=1.0]{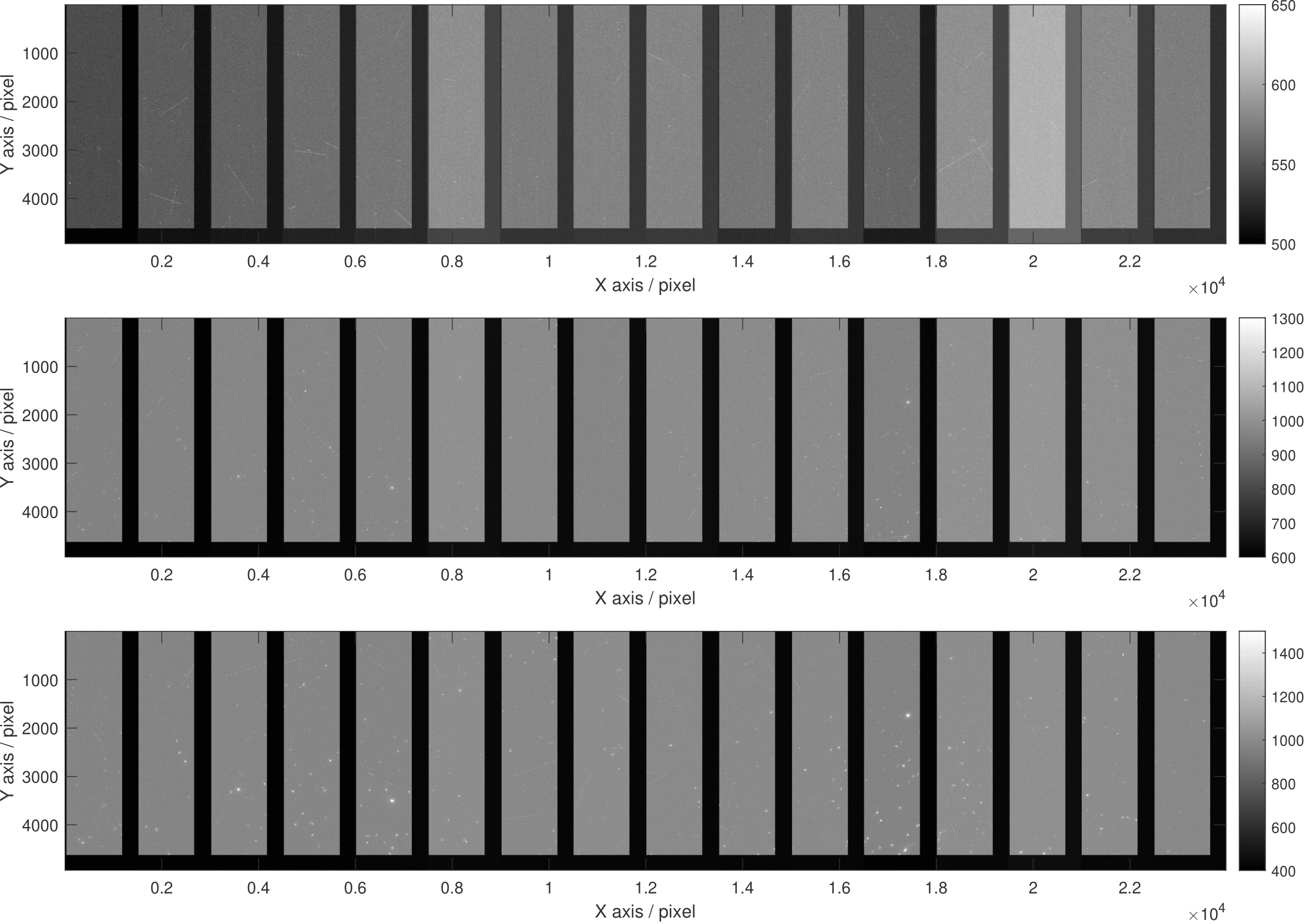}
   \caption{A demonstration of level 0 scientific data images obtained from the MCI's three-channel ultra-deep field image simulation. The ultraviolet channel (top) . The visible channel (middle). The near-infrared channel (bottom).}
   \label{EXDF_img}
\end{figure}

Figure~\ref{EXDF_img} demonstrates level 0 scientific data images from the MCI's three-channel galaxy simulation. This simulation uses the galaxy catalog described in Section~\ref{Image simulation of galaxies}. The three channels employ 'CBU', 'CBI', and 'CBV' filters with a 900-second single exposure time. Cosmic rays are clearly visible in the simulated images, along with faint galaxy signatures. However, most galaxies are sufficiently faint that their signals are overwhelmed by noise in single exposures. To detect these fainter sources, a multi-exposure approach is required. Subsequent processing applies image stacking and deconvolution algorithms to enhance the signal-to-noise ratio (SNR), as presented in the following section.

\subsection{Photometric calibration simulation results}
To ensure the observational accuracy of CSST-MCI, the open cluster NGC 2477 was selected as one of the calibration star fields. It is used as a case study to assess the performance of the MCI data processing pipeline in photometric calibration. This cluster offers distinct advantages: high stellar density, abundant Gaia XP spectral data within the MCI’s field of view, and a relatively uniform stellar distribution, all of which make it an ideal reference for calibrating MCI’s photometric and astrometric precision. The main simulation and data processing workflow are as follows.

\begin{figure}[htpb]
   \centering
 \includegraphics[width=\textwidth, angle=0, scale=0.8]{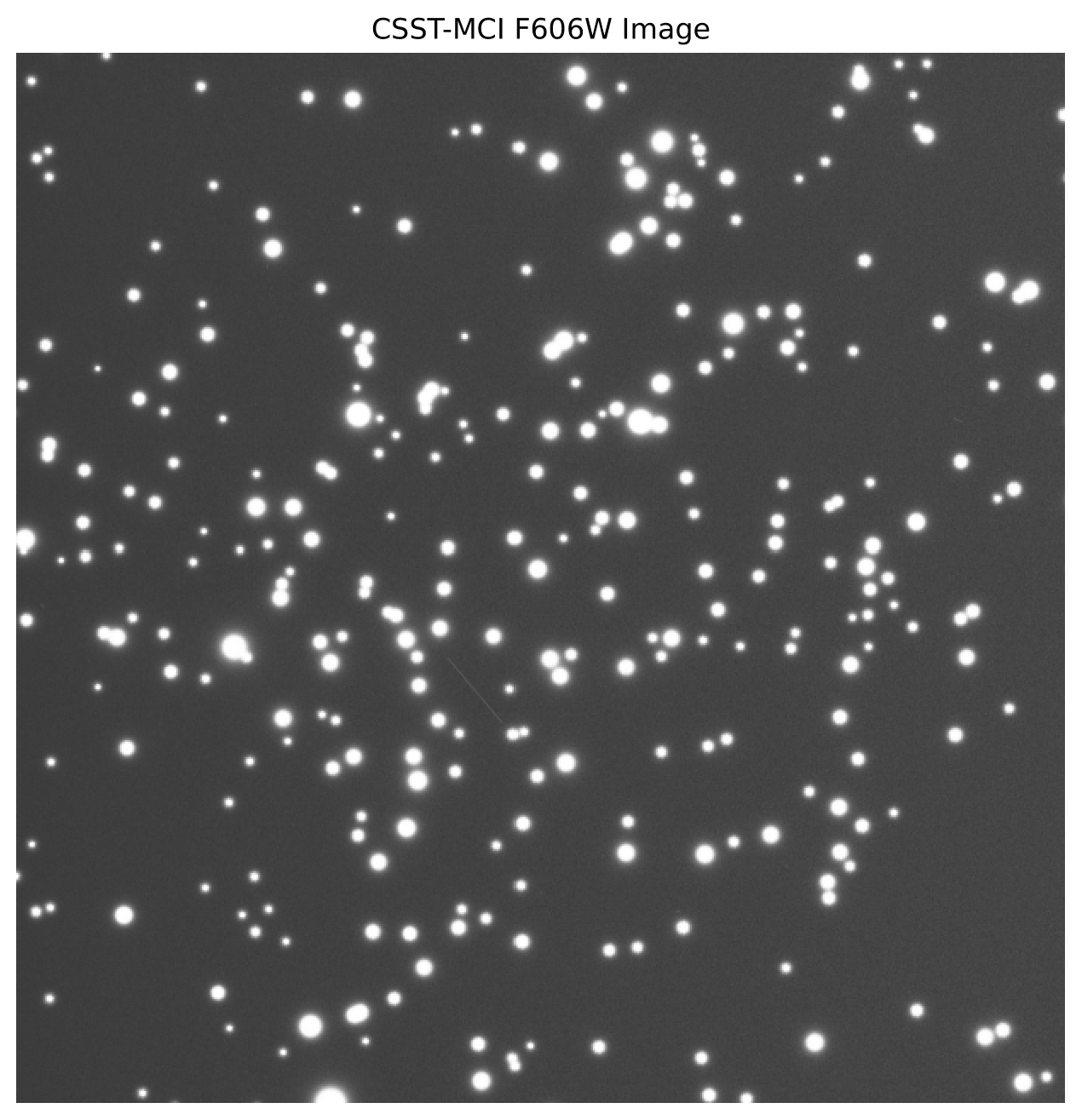}
   \caption{CSST-MCI simulated image of NGC2477 in the F606W band (level 1 scientific data image).}
   \label{CSST_F606W_img}
\end{figure}

\begin{enumerate}
    \item \textbf{Input Data Selection:} The Gaia catalog of NGC 2477 serves as the foundational dataset, providing astrometric information (positions, proper motions) and photometric data (magnitudes) for stars in the cluster. Additionally, we selected over 300 stars with G-band magnitudes greater than 15 within the MCI’s field of view, using their Gaia XP sample SED data as spectral templates.

    \item \textbf{Dense Star Field Image Generation:} Based on these spectral templates, we conducted simulations in the F606W band to generate high-density star field images. This process involves the precise modeling of stellar spectral characteristics to realistically reproduce stellar radiative energy across different wavelengths, providing input scenarios close to real observations for subsequent data processing. Figure~\ref{CSST_F606W_img} shows the simulation result of NGC2477 in the F606W band (level 1 scientific data image). The simulation does not incorporate all instrumental effects; instead, it primarily takes into account fundamental factors such as detector noise, sky background noise, and cosmic ray events.

    \item \textbf{Application of the CSST-MCI 1D Data Processing Pipeline:} The simulated star field images were fed into the CSST-MCI 1D data processing pipeline for preprocessing. This step involves standard procedures such as bias subtraction, flat-field correction, and source extraction, ultimately yielding the instrumental magnitudes of each star in the dense star field. These results serve as a critical benchmark for evaluating MCI performance in crowded stellar environments.

    \item \textbf{Flux Calibration via Bayesian Inference:} For stars in the simulated field, their Gaia XP spectral flux ($f_{\lambda}$) and filter photometry counts ($C_P$, linked to instrumental magnitudes) serve as input data. Within the Bayesian framework, the posterior probability distribution function (PDF) of the calibration parameters in this photometric system can be expressed as:
\begin{equation}	
\label{eq:1}
 P(f_{\lambda}(P), C_P | T_P, P_{\lambda}) \propto \mathscr{L} ( T_P,P_{\lambda} |f_{\lambda}, C_P) \centerdot \pi ( T_P,P_{\lambda})
\end{equation}
where $\pi ( T_P,P_{\lambda})$ is the prior probability of the PDF, and the likelihood function can be expressed as:
\begin{equation}	
\label{eq:2}
\begin{aligned}
 \mathscr{L} ( T_P |f_{\lambda}, C_P) = \prod_i \mathrm{exp} [\frac{f_{\lambda}(P)(i)/T_P-C_P(i)}{\sigma(i)}]^2
 \end{aligned}
\end{equation}
where $f_{\lambda}(P)(i)$, $C_P(i)$ and $\sigma(i)$ represent the energy spectrum distribution of the i-th star after convolution with $P_{\lambda}$, the ADU counts of the photometric system, and the ADU counts's error, respectively.

\end{enumerate}

We employed the Markov Chain Monte Carlo (MCMC) method with the public package $emcee$ \citep{2013PASP..125..306F} to sample the flux conversion coefficient $T_P$ of the photometric system and the transmittance $P_{\lambda}$ at each wavelength to obtain their posterior probability distributions. We subsequently used the Kernel Density Estimation (KDE) method to determine the maximum probability values of the parameters within the posterior probability distributions, thereby obtaining the calibration parameters $T_P$ and $P_{\lambda}$ under this photometric system.

\begin{figure}[htpb]
   \centering
 \includegraphics[width=\textwidth, angle=0, scale=0.7]{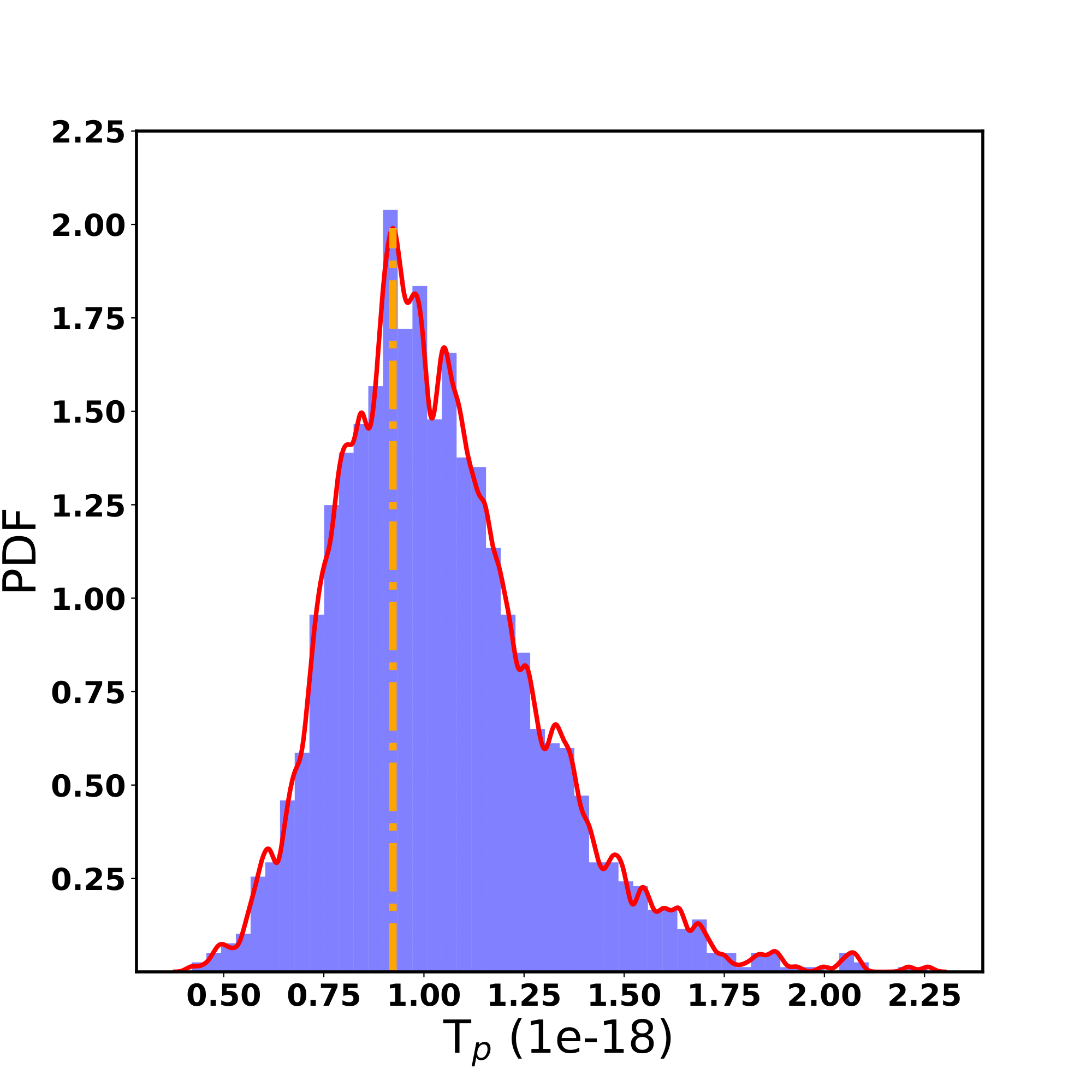}
   \caption{Probability density distribution of the flux conversion coefficient obtained from the MCI simulation experiment. The red solid line represents the probability density distribution obtained via the KDE method, and the yellow dotted-dashed line marks the $T_P$ value corresponding to the maximum probability.}
   \label{photflam}
\end{figure}

\begin{figure}[htpb]
   \centering
 \includegraphics[width=\textwidth, angle=0, scale=0.7]{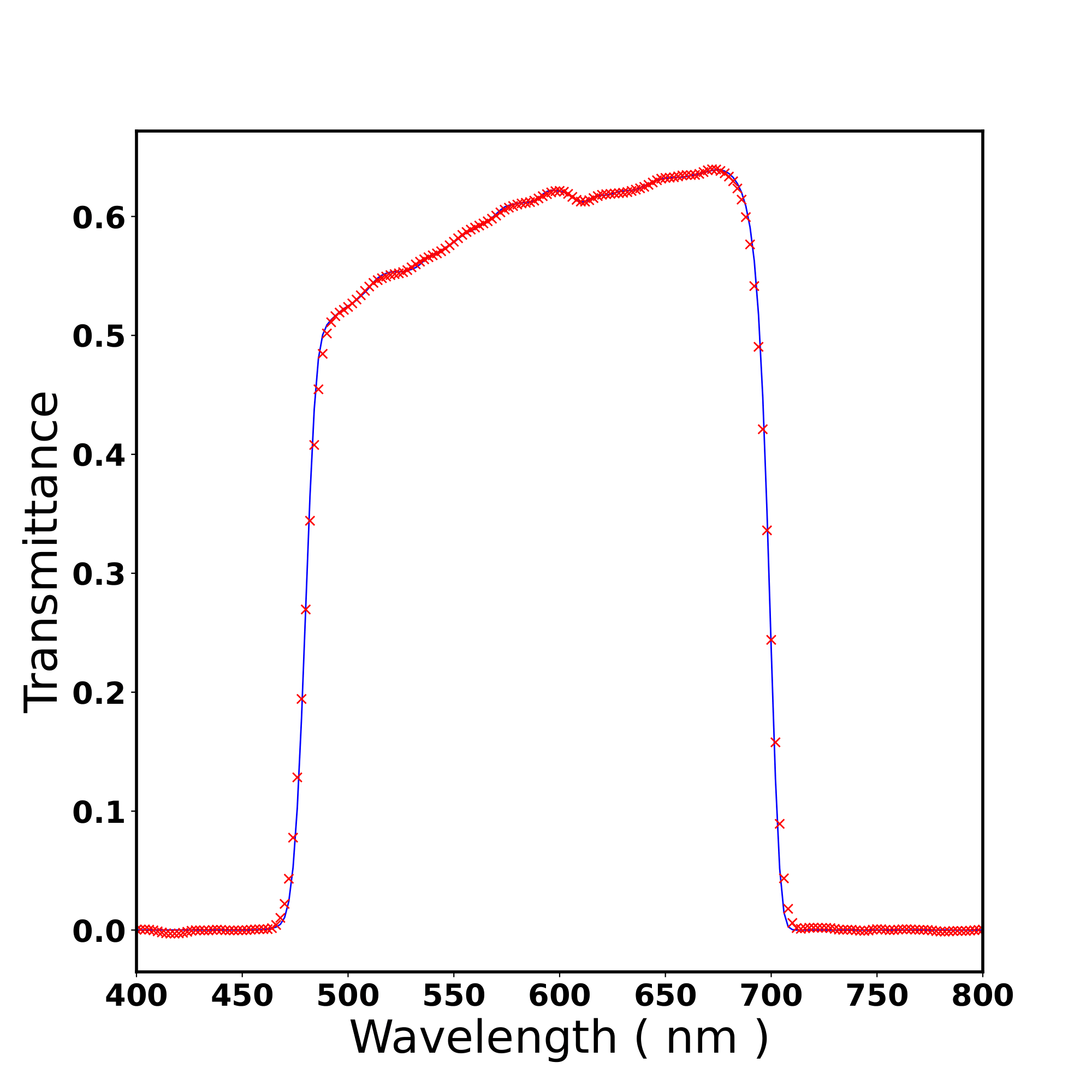}
   \caption{Transmittance curve of the F606W filter. The blue solid line represents the input, and the red crosses represent the transmittance curves of the photometric system obtained through Bayesian inference based on simulation data. The MCI flux calibration reproduced the transmittance curve well.}
   \label{transmission}
\end{figure}

Figure~\ref{photflam} and Figure~\ref{transmission} show the Bayesian inference results of two key parameters for flux calibration, $T_P$ and $P_{\lambda}$, respectively. Based on the obtained flux conversion coefficient $T_P$ , the calibration zero point of the STMAG magnitude system can be calculated according to the formula below \citep{2005PASP..117.1049S}.

\begin{equation}	
\label{eq:3}
\begin{aligned}
Zpt=-2.5\log[T_p] - 21.10
 \end{aligned}
\end{equation}

This workflow leverages NGC 2477’s dense stellar population and rich Gaia XP data to systematically validate MCI’s photometric accuracy, ensuring reliable measurements in complex astronomical environments.

Notably, the MCI is designed for flux calibration owing to its superior photometric precision. Specifically, to support the main survey in achieving high-accuracy flux calibration, the MCI is designed to match the high-precision photometric capabilities of the CSST. This includes stringent requirements on stray light suppression, high-speed shutter, focus adjustment, and the quality of optical components and detectors. Unlike the main survey camera, which is optimized for wide-field survey performance, the MCI is optimized for precision measurements. The accuracy of flux calibration primarily hinges on photometric precision, which is typically achieved through two main methods. The first is aperture photometry, where the aperture size is determined based on the growth curve. In this approach, variations in the PSF shape within a given aperture have a minimal impact on photometric accuracy and are generally negligible. The second method is PSF photometry, which offers superior precision for dense stellar fields, as its performance is directly tied to the accuracy of the PSF model. The fidelity of PSF modeling, in turn, depends on image sampling—smaller PSFs lead to poorer sampling, which can hinder accurate modeling, particularly in undersampled regimes \citep{2000PASP}. In the main survey, the pixel scale is $0.074^{''}$ with an average EE80 radius of approximately $0.13^{''}$. By comparison, the MCI has a finer pixel scale of $0.05^{''}$ and a larger EE80 radius of approximately $0.18^{''}$. This means that the MCI PSF is better sampled than that of the main survey, enabling more accurate PSF modeling and potentially more precise flux calibration. These high-precision flux standards provided by the MCI can then be used to calibrate the main survey, enabling robust and accurate flux calibration across the full dataset.

\subsection{Simulation results of the Drizzle coaddition algorithm}

\begin{figure}[htpb]
   \centering
 \includegraphics[width=\textwidth, angle=0, scale=1.0]{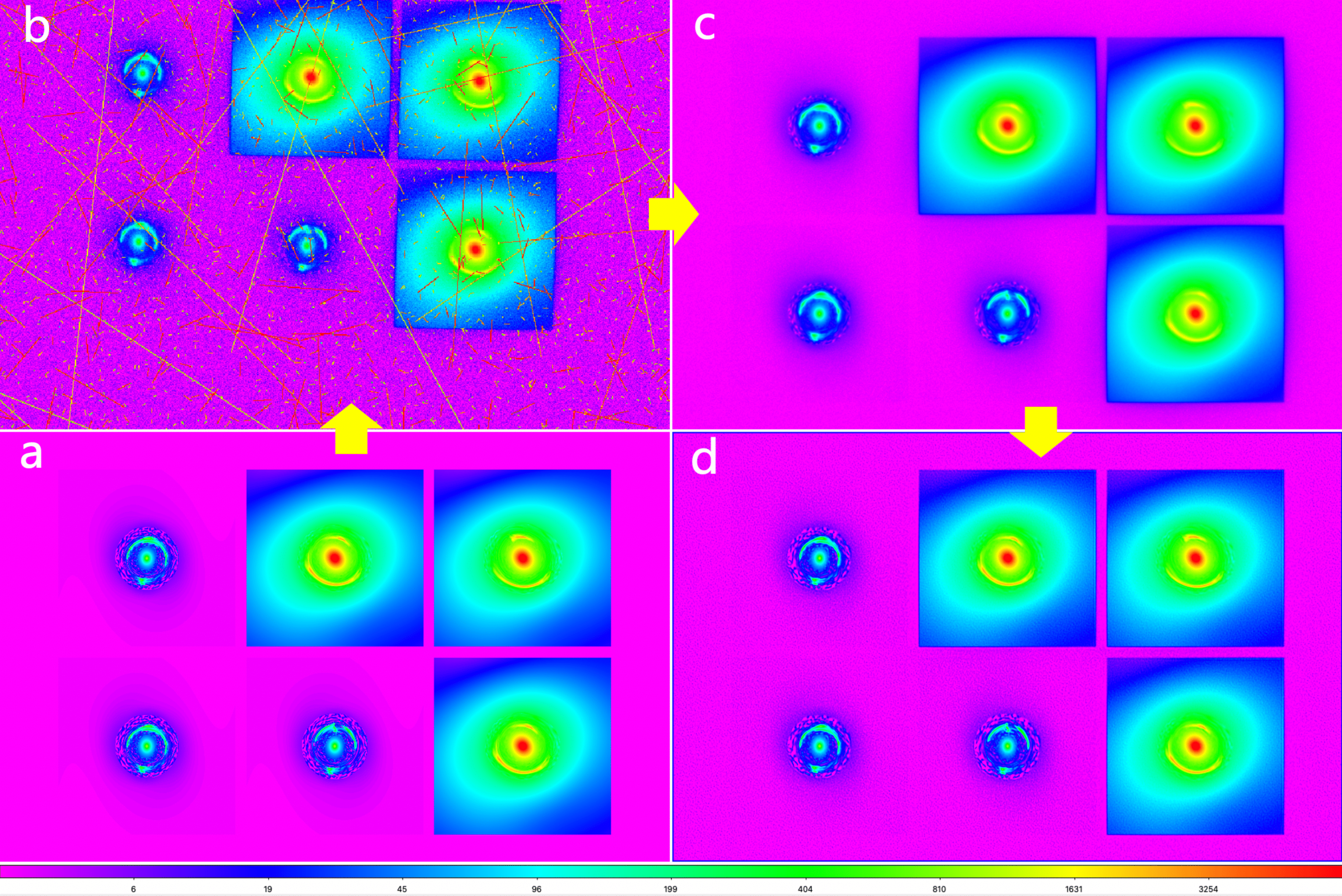}
   \caption{Processing effects of mocked multiple-exposure strong gravitational lensing images after passing through the CSST-MCI pipeline. Panel (a) is the input ground truth. Panel (b) is a mocked MCI image from a single exposure, which downsamples panel (a) by a factor of 2 and adds observational effects including the PSF and cosmic rays. Panel (c) shows the Drizzle coaddition result of 100 exposures like panel (b). Panel (d) shows the coaddition result using the UPDC algorithm on the same 100 exposures.}
   \label{mocked_1}
\end{figure}

We evaluate the performance of the MCI data processing pipeline for ultra-deep field via strong gravitational lensing as a case study. We apply the UPDC algorithm\citep{Wang+2022}, which was specifically developed for CSST-MCI, to coadd 100 mock exposures in the g-band. The results are presented in Figure~\ref{mocked_1}. Panel (a) is the input ground truth. Panel (b) is a single-exposure simulated MCI image. Panel (c) is the Drizzle coaddition result of 100 exposures, similar to panel (b). Panel (d) is the coaddition result using the UPDC algorithm on the same 100 exposures.

It is evident that the UPDC algorithm's output is closer to the input ground truth (panel a). A comparison of the Drizzle result with the UPDC result reveals that the latter reveals more spatial details of the lensed image substructures. This enhanced detail proves advantageous in obtaining a clearer spatial morphology of the background galaxy during subsequent modeling.

\begin{figure}[htpb]
   \centering
 \includegraphics[width=\textwidth, angle=0, scale=1.0]{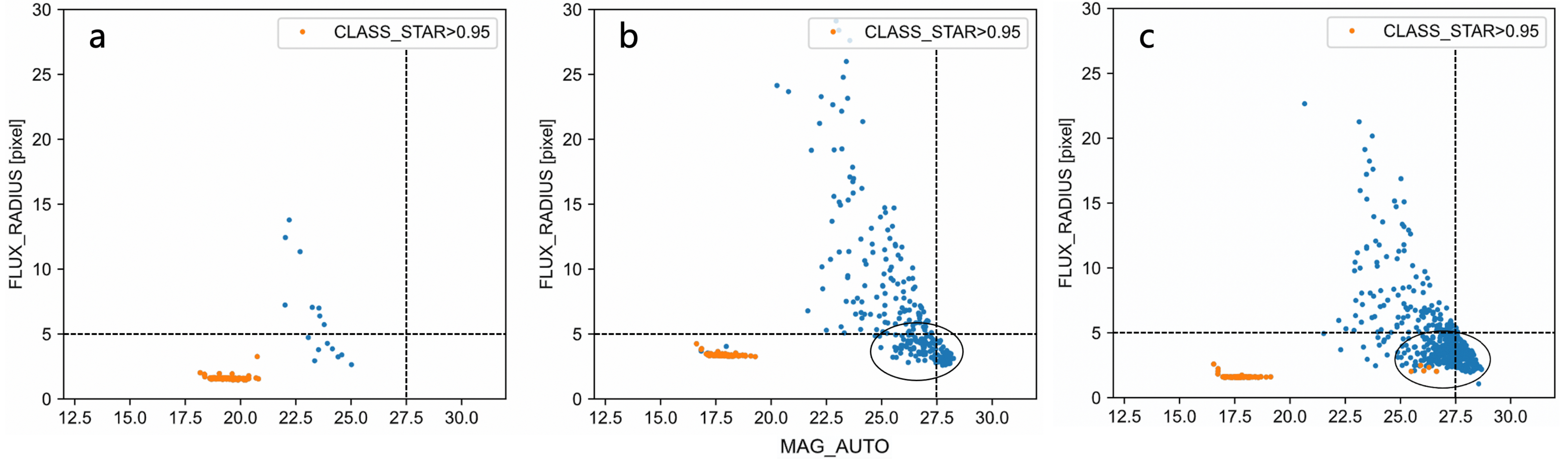}
   \caption{Improvement in detection efficiency for faint sources in CSST-MCI mock observations through multi-exposure image coaddition. Panel (a) shows the "source magnitude vs. flux radius" diagram obtained by source detection using SExtractor++ software on a mock sky region for the MCI C1 channel. The orange points represent point sources, and the blue points represent extended sources (detection threshold: $6\sigma$). Panel (b) displays faint sources detectable after coadding 500 mock exposure images using the traditional Drizzle algorithm. Compared to panel (a), the image is upsampled by a factor of 1, resulting in a doubling of the point source radius. Panel (c) shows the coaddition result using the UPDC algorithm on the same mock exposures. Using the same detection threshold, we detect more and fainter sources than those in the Drizzle results.}
   \label{mocked_2}
\end{figure}

Given that detecting high-redshift faint targets is a key scientific objective of CSST-MCI, Figure~\ref{mocked_2} illustrates the enhanced detection efficiency for faint sources in CSST-MCI mock observations through multi-exposure image coaddition. Panel (a) presents the "source magnitude vs. flux radius" diagram from source detection using SExtractor++ software on a mock sky region for the C1 channel of the MCI. The orange points denote point sources, whereas the blue points represent extended sources (detection threshold: $6\sigma$). Panel (b) shows faint sources detectable after coadding 500 mock exposures using the traditional Drizzle algorithm. Compared to panel (a), the image is upsampled by a factor of 2, leading to doubled point source radii. Panel (c) demonstrates the coaddition result using UPDC on the same mock exposures. Under identical detection thresholds, the UPDC detects a significantly greater number of faint sources compared to the Drizzle algorithm.

Moreover, as UPDC deconvolves the PSF and applies anti-aliasing for undersampled exposures, the FWHM of both faint and point sources is significantly reduced. The detectable faint source magnitude limit improves by roughly 3.1 magnitudes after Drizzle coaddition compared to single exposures, aligning with the theoretical expectation of 3.37 magnitudes. With UPDC coaddition, this limit is further enhanced by an additional 0.2 to 0.3 magnitudes. Comparable enhancements have also been observed in JWST image coaddition\citep{Wang+2025}.

\section{Conclusions}

We present a comprehensive overview of scientific image simulations for the CSST-MCI. The optical models of both the CSST and MCI systems are described, incorporating wavefront aberrations from CSST engineering simulations as the basis for multi-wavelength PSF calculation and interpolation. The underlying mathematical principles are detailed, enabling the generation of star and galaxy images with rendered PSFs. The simulations account for numerous instrumental effects: shutter operations, cosmic rays, astrometric distortions, sky background noise, stray light, and various sensor artifacts. Developed over several years, the simulation software produces data products that have been rigorously tested and iteratively refined within the MCI data processing pipelines. These validated simulations serve as a foundation for the team's scientific research. Current simulation parameters are derived from literature and manufacturer data, as direct MCI detector measurements are unavailable. These values represent our best estimates while maintaining physical plausibility, and they serve as a critical foundation for preliminary studies. However, this approach introduces inherent uncertainties—particularly regarding potential discrepancies between the assumed and actual characteristics of the MCI detector—reflecting the current stage of development. At present, laboratory test systems are unable to fully characterize certain instrumental effects, such as the full-field flat response, the brighter-fatter effect, and charge diffusion. As a result, we have adopted provisional parameters from detectors with similar architectures, such as those used in LSST. To systematically address these limitations, we have established a three-phase refinement strategy: (1) incorporating laboratory-measured parameters into upcoming versions of the simulation software; (2) conducting comprehensive pre-launch calibration; and (3) performing post-launch analyses using on-orbit observational data. This iterative approach ensures the simulation framework continues to evolve, providing increasingly accurate support for mission planning and scientific preparatory work as more definitive measurements become available.

\begin{acknowledgements}
HYS acknowledges the support from the Ministry of Science and Technology of China (grant No. 2020SKA0110100), the NSFC of China under No.11973070, No.11873078, Key Research Program of Frontier Sciences, CAS, Grant No. ZDBS-LY-7013 and Program of Shanghai Academic/Technology Research Leader. We acknowledge the support from the science research grants from the China Manned Space Project with NO. CMS-CSST-2021-A01, CMS-CSST-2021-A04.  Jing Zhong would like to acknowledge the science research grants from the China Manned Space Project with NO. CMS-CSST-2025-A19, the Science and Technology Commission of Shanghai Municipality (Grant No.22dz1202400), and Sponsored by the Program of Shanghai Academic/Technology Research Leader.
\end{acknowledgements}

\bibliographystyle{raa}
\bibliography{bibtex}

\label{lastpage}
\appendix
\section{Instrument effects and function module switches}
The instrument effect switches and function module switches that can be set by the user are as follows.

\# apply multiplicative flat field or not

$\bullet$ flatfieldM = yes

\# add CCD dark current or not

$\bullet$ darknoise = yes

\# add skylight background and stray light or not

$\bullet$ skyback = yes

\# apply cosmetic defects to CCD or not

$\bullet$ cosmetics = yes

\# apply radiation damage effect or not

$\bullet$ radiationDamage = yes

\# add cosmic rays or not  

$\bullet$ cosmicRays = yes

\# apply bleeding effect or not

$\bullet$ bleeding = yes

\# apply non-linearity effect or not

$\bullet$ nonlinearity = yes

\# apply readout noise or not

$\bullet$ readoutnoise = yes

\# save cosmicrays map or not

$\bullet$ save\_cosmicrays = no

\# apply prescan and overscan effect or not

$\bullet$ overscans = yes

\# apply astrometric effect or not

$\bullet$ TianceEffect = yes
                                              
\# convert a floating point image array to an integer array or not

$\bullet$ intscale = yes

\# apply optical ghost effect or not

$\bullet$ ghosts = yes

\# apply shutter effect or not

$\bullet$ shutterEffect = yes
                      
\# apply pixel-to-pixel non-uniformity effect or not  

$\bullet$ PRNUeffect = yes 

\# apply bright fatter effect or not

$\bullet$ appFatt = yes

\# apply sky rotation and shift effect or not

$\bullet$ sky\_shift\_rot = yes

\# apply optical field distortion effect or not

$\bullet$ distortion = yes

\# simulate star image or not

$\bullet$ sim\_star = yes

\# simulate galaxy image or not

$\bullet$ sim\_galaxy = yes

\# save PSF data or not

$\bullet$ save\_starpsf = yes

\# apply gravitational lensing  or not

$\bullet$ lensing = yes

\end{document}